\documentstyle[12pt]{article}
\input epsf
\epsfverbosetrue
\newlength{\dinwidth} 
\newlength{\dinmargin}
\begin{document}
\newcommand{\ra}{\rightarrow}
\newcommand{\as}{\mbox{$\alpha_{\displaystyle  s}$}}
\newcommand{\mxtwo}{\mbox{$M^2_x$}}
\newcommand{\Qtwo}{\mbox{$Q^2$}}
\newcommand{\thetah}{\mbox{$\theta_H$}}
\newcommand{\etamax}{\mbox{$\eta_{\rm max}$}}
\newcommand{\betan}{\mbox{$\beta\;$}}
\newcommand{\xbj}{\mbox{$\rm x_{Bj}$}}

\newcommand {\pom}  {I\hspace{-0.2em}P}
\newcommand {\xpom} {\mbox{$x_{_{\pom}}$}}
\newcommand {\apom} {\mbox{$\alpha_{_{\pom}}$}}
\newcommand {\aprime} {\mbox{$\alpha^\prime_{_{\pom}}$}}
\newcommand {\xpomp}[1] {\mbox{$x^{#1}_{_{\pom}}\;$}}
\newcommand {\xpoma} {\mbox{$(1/x_{_{\pom}})^{\Large a}\;$}}
\newcommand {\xl} {\mbox{$x_L$}}

\newcommand {\mxlps} {\mbox{$ M_{X}^{LPS} $}}
\newcommand {\mxlpstwo} {\mbox{$(M_{X}^{LPS})^2$}}
\newcommand {\mxta} {\mbox{$M_{X~CAL}$}}
\newcommand{\mxz} {\mbox{$M_X^{meas}$}} 
\newcommand{\mev}{\mbox{$\rm MeV$}}
\newcommand{\gev}{\mbox{$\rm GeV$}}
\newcommand{\gevtwo}{\mbox{$\rm GeV^2$}}
\newcommand{\gevmtwo}{\mbox{$\rm GeV^{-2}$}}
\newcommand{\cm}{\mbox{$\rm cm$}}
\newcommand{\cmtwo}{\mbox{$\rm cm^2$}}
\newcommand{\nbi}{\mbox{$\rm nb^{-1}$}}
\newcommand{\pbi}{\mbox{$\rm pb^{-1}$}}
\newcommand{\sleq} {\raisebox{-.6ex}{${\textstyle\stackrel{<}{\sim}}$}}
\newcommand{\sgeq} {\raisebox{-.6ex}{${\textstyle\stackrel{>}{\sim}}$}}
\newcommand{\gp}{\mbox{$\gamma^* p$}}
\newcommand{\ftwod}{\mbox{$F_2^{D(3)}$}}
\newcommand{\ftwodfour}{\mbox{$F_2^{D(4)}$}}
%
%
%
%
%
%
%
%
\newcommand{\ftwodfoura}{\mbox{1.00}}
\newcommand{\ftwodfouraerr}{\mbox{$\ftwodfoura 
     \pm 0.09 ~{\rm (stat.)}^{+0.11}_{-0.05} {\rm (syst.)}$}}

\newcommand{\ftwodfouralpha}{\mbox{1.00}}
\newcommand{\ftwodfouralphaerr}{\mbox{$\ftwodfouralpha 
     \pm 0.045 ~{\rm (stat.)}^{+0.08}_{-0.025} {\rm (syst.)}$}}

\newcommand{\ftwodfouralphaz}{\mbox{1.04}}
\newcommand{\ftwodfouralphazerr}{\mbox{$\ftwodfouralphaz 
     \pm 0.045 ~{\rm (stat.)}^{+0.08}_{-0.025} {\rm (syst.)}$}}

\newcommand{\ftwodthreea}{\mbox{1.01}}
\newcommand{\ftwodthreeaerr}{\mbox{$\ftwodthreea 
     \pm 0.10 ~{\rm (stat.)}^{+0.11}_{-0.06} {\rm (syst.)}$}}
\newcommand{\ftwodfourtmax}{\mbox{0.4}}

\newcommand{\tslopeerr} {\mbox{$7.2\pm 1.1 {\rm (stat.)}^{+0.7}_{-0.9}
{\rm (syst.)}~\gevmtwo$}}

\begin{flushleft}
\tt DESY 97-184
\end{flushleft}

\vspace{1cm}

\begin{center}
\LARGE{ 
{\bf Measurement of the Diffractive Structure \\ 
Function {\boldmath $F_2^{D(4)}$} at HERA }} 
\end{center}

\vspace{2cm}
\begin{center}
\Large{ZEUS Collaboration}
\end{center}
\date{ }

%
\vspace{2cm}


\vspace{1cm}

\begin{abstract}

This paper presents the first analysis of diffractive photon dissociation
events in deep inelastic positron-proton scattering at HERA in which  the proton
in the final state is detected and its momentum measured.
The events are selected by requiring 
a scattered proton 
in the ZEUS leading proton spectrometer (LPS)
with $\xl>0.97$,
where $\xl$ is the fraction of the incoming proton beam momentum carried by the
scattered proton.
The use of the LPS significantly reduces the contamination from
events with diffractive dissociation of the proton into low mass states
and allows a direct measurement of $t$, the square of the four-momentum exchanged 
at the proton vertex.
The dependence of the cross section on $t$ is measured in the interval
$0.073<|t|<0.4$~$\gevtwo$ and is found to be described by  
an exponential shape with the slope parameter 
$b=\tslopeerr$. 
The diffractive structure function $\ftwodfour$ is presented 
as a function of $\xpom \simeq 1-\xl$ and  $\beta$,
the momentum fraction of the struck quark with respect to $\xpom$,
and averaged over the $t$ interval
$0.073<|t|<\ftwodfourtmax$~$\gevtwo$ and
the photon virtuality range $5<Q^2<20~\gevtwo$. 
In the kinematic  range 
$4 \times 10^{-4} < \xpom < 0.03$ and $0.015<\beta<0.5$,
the $\xpom$ dependence of $\ftwodfour$ is fitted
with a form
$\xpoma$, yielding $a= \ftwodfouraerr$. 
Upon integration over $t$, the structure
function $\ftwod$ is determined in a kinematic range 
extending 
to higher $\xpom$ and lower $\beta$ compared to our
previous analysis;
the  results are discussed within the framework of Regge theory.

\end{abstract}

%
%
\setcounter{page}{0}
\thispagestyle{empty}
\pagenumbering{Roman}                                                           
\def\3{\ss}
                   
\newpage
%
%
%
%
                                                   %
\begin{center}                                                                                     
{                      \Large  The ZEUS Collaboration              }                               
\end{center}                                                                                       
  J.~Breitweg,                                                                                     
  M.~Derrick,                                                                                      
  D.~Krakauer,                                                                                     
  S.~Magill,                                                                                       
  D.~Mikunas,                                                                                      
  B.~Musgrave,                                                                                     
  J.~Repond,                                                                                       
  R.~Stanek,                                                                                       
  R.L.~Talaga,                                                                                     
  R.~Yoshida,                                                                                      
  H.~Zhang  \\                                                                                     
 {\it Argonne National Laboratory, Argonne, IL, USA}~$^{p}$
% ...
\newpage                                                                                           
% ...
$^{  23}$ now at ATLAS Collaboration, Univ. of Munich \\                                           
% ...
$^{  32}$ partially supported by the Foundation for German-Russian Collaboration                   
DFG-RFBR \\ \hspace*{3.5mm} (grant no. 436 RUS 113/248/3 and no. 436 RUS 113/248/2)\\              
% ...
                                                           %
                                                           %
\newpage   
                                                           %
                                                           %
\begin{tabular}[h]{rp{14cm}}                                                                       
$^{a}$ &  supported by the Natural Sciences and Engineering Research                               
          Council of Canada (NSERC)  \\                                                            
$^{b}$ &  supported by the FCAR of Qu\'ebec, Canada  \\                                            
$^{c}$ &  supported by the German Federal Ministry for Education and                               
          Science, Research and Technology (BMBF), under contract                                  
          numbers 057BN19P, 057FR19P, 057HH19P, 057HH29P, 057SI75I \\                              
$^{d}$ &  supported by the MINERVA Gesellschaft f\"ur Forschung GmbH,                              
          the German Israeli Foundation, and the U.S.-Israel Binational                            
          Science Foundation \\                                                                    
$^{e}$ &  supported by the German Israeli Foundation, and                                          
          by the Israel Science Foundation                                                         
  \\                                                                                               
$^{f}$ &  supported by the Italian National Institute for Nuclear Physics                          
          (INFN) \\                                                                                
$^{g}$ &  supported by the Japanese Ministry of Education, Science and                             
          Culture (the Monbusho) and its grants for Scientific Research \\                         
$^{h}$ &  supported by the Korean Ministry of Education and Korea Science                          
          and Engineering Foundation  \\                                                           
$^{i}$ &  supported by the Netherlands Foundation for Research on                                  
          Matter (FOM) \\                                                                          
$^{j}$ &  supported by the Polish State Committee for Scientific                                   
          Research, grant No.~115/E-343/SPUB/P03/002/97, 2P03B10512,                               
          2P03B10612, 2P03B14212, 2P03B10412 \\                                                    
$^{k}$ &  supported by the Polish State Committee for Scientific                                   
          Research (grant No. 2P03B08308) and Foundation for                                       
          Polish-German Collaboration  \\                                                          
$^{l}$ &  partially supported by the German Federal Ministry for                                   
          Education and Science, Research and Technology (BMBF)  \\                                
$^{m}$ &  supported by the Fund for Fundamental Research of Russian Ministry                       
          for Science and Edu\-cation and by the German Federal Ministry for                       
          Education and Science, Research and Technology (BMBF) \\                                 
$^{n}$ &  supported by the Spanish Ministry of Education                                           
          and Science through funds provided by CICYT \\                                           
$^{o}$ &  supported by the Particle Physics and                                                    
          Astronomy Research Council \\                                                            
$^{p}$ &  supported by the US Department of Energy \\                                              
$^{q}$ &  supported by the US National Science Foundation \\                                       
\end{tabular}                                                                                      
                                                           %
                                                           %

\newpage
\pagenumbering{arabic}
\setcounter{page}{1}
\normalsize

\section{Introduction}

A distinct class of events in
deep inelastic neutral current positron-proton scattering (DIS) 
is experimentally characterized by the proton emerging 
intact (or excited into a low-mass state) and 
well separated in rapidity from the state produced by the
dissociation of the virtual photon \cite{zeuslrg,h1lrg}.
These large rapidity gap events
 can be interpreted as being due to diffractive interactions mediated 
by the  exchange of a colourless
object with the vacuum quantum numbers,
generically called the pomeron.

In diffractive single dissociation DIS events at HERA, 
$ep \rightarrow e X p$, 
the virtual photon dissociates into 
a hadronic system of mass $M_X$, while the proton remains intact; 
for small $|t|$ ($|t| \sleq 1~\gevtwo$), where $t$ is the square of the
four-momentum transfer at the proton vertex, the scattered proton 
escapes through the beam pipe  without being detected in the central apparatus. 
Therefore
diffractive interactions have been studied so far in the H1 and ZEUS experiments
either by requiring
a large rapidity gap \cite{h1f2diff,zeusf2diff,h1f2diff94}
or exploiting the different behaviour of the $\ln(M_X^2)$ distribution for
diffractive and non-diffractive events 
\cite{kowalski}. 
In both approaches, in the ZEUS results, the hadronic mass $M_X$
was limited to values below $20 ~\gev$.

This paper presents the first analysis of diffractive DIS events
in which the proton in the final state is detected
and its momentum is measured. The 
measurement was performed with the ZEUS detector at HERA, 
using the  leading proton spectrometer (LPS) which detects
the scattered
proton at small angles ($\sleq 1$ mrad) with respect to the incoming proton beam
and measures its momentum.
The events were selected 
by requiring that the scattered proton carry
a fraction of the incident proton beam momentum, $x_L$, greater than 0.97, 
 a region where pomeron exchange dominates. 
The use of the LPS allows a direct measurement of $t$, and extends
the measurement to higher values of $M_X$ than in our previous analyses.

\section{The diffractive structure function {\boldmath $\ftwodfour$}}

The dependence of the total 
hadron-hadron and photon-hadron 
cross sections on the centre of mass (c.m.)  energy 
is related in the Regge approach to 
the pomeron trajectory. A fit to the hadron-hadron data
\cite{dl} yielded a universal pomeron trajectory 
$\apom(t) = \apom(0) + \aprime t$,
with $\apom(0) \simeq 1.08$   
(the `soft' pomeron).
The dependence of the elastic hadron-hadron cross section on $t$ 
 is well described by an exponential distribution 
at small $|t|$,
with a slope that increases with the c.m. energy (shrinkage), leading to
$\aprime \simeq 0.25 ~\gevmtwo$.
Regge theory can also be used to describe inclusive diffractive
dissociation~\cite{mueller}.
These processes have been studied in hadron-hadron
(see e.g.~\cite{goulianos}-\cite{cdf}), 
photon-hadron (\cite{e612}-\cite{zeusphp}) and 
DIS (\cite{h1f2diff}-\cite{kowalski}) interactions. 

For diffractive single dissociation in DIS, 
$e^+(k) p(P) \rightarrow e^+(k^\prime) X p(P^\prime) $, the cross section can
be expressed in terms of the diffractive structure function
$\ftwodfour$:
\begin{equation}
  \frac{d^4\sigma_{diff}}{d\beta dQ^2 d\xpom dt} = \frac{2 \pi
    \alpha^2}{\beta Q^4} \; (1+(1-y)^2) \;
  F_2^{D(4)}(\beta,Q^2,\xpom,t),
\label{first}
\end{equation}
where $\alpha$ is the electromagnetic coupling constant and 
the contributions of the longitudinal structure function and of
$Z^0$ exchange have been neglected. Note that, with this definition,
$\ftwodfour$ has dimensions of $\gevmtwo$. Integrating over $t$, we can define
a structure function $\ftwod$:
\begin{equation}
  \frac{d^3\sigma_{diff}}{d\beta dQ^2 d\xpom} = \frac{2 \pi
    \alpha^2}{\beta Q^4} \; (1+(1-y)^2) \;
  F_2^{D(3)}(\beta,Q^2,\xpom).
\label{first2}
\end{equation}

The relevant kinematic variables are defined 
as follows. Deep inelastic scattering events are described by 
$Q^2= -q^2=-(k-k^\prime)^2$, the negative of the squared four-momentum transfer
carried by the virtual photon; by the Bjorken variable $x= \frac{Q^2}{2P \cdot q}$;
and by $y=Q^2/xs$, the fractional
energy transferred to the proton in its rest frame, where $\sqrt s=300~\gev$ is the $ep$ c.m.
 energy. The c.m. energy of the virtual photon-proton ($\gamma^*p$) system is
$W \simeq \sqrt{Q^2(1/x-1)}$.

Additional variables are required to describe diffractive scattering:
\begin{equation}
t=(P-P^\prime)^2,~~ 
\xpom = \frac{(P-P^\prime)\cdot q}{P \cdot q},~~ 
\beta = {\frac{Q^2}{2(P-P^\prime) \cdot q}}={\frac{x}{\xpom}},
\end{equation}
where  $\xpom$ is the momentum fraction of the proton carried by the pomeron.
In models where the pomeron has a partonic structure (see e.g. \cite{ingsch}),
$\beta$ is the momentum fraction of the struck quark within the pomeron.

Assuming factorization, as in the model of Ingelman and Schlein \cite{ingsch},
the structure function $\ftwodfour$ is factorized into a pomeron
flux, depending on $\xpom$ and $t$, and a pomeron structure function,
which depends on $\beta$ and $Q^2$:
\begin{equation}
\ftwodfour = f_{\pom}(\xpom,t) \cdot F_2^{\pom}(\beta,Q^2).
\end{equation}
In Regge theory the $\xpom$ dependence of the flux
can be expressed as $(1/\xpom)^{2 \apom(t)-1}$.

In this paper, we present a measurement of the structure function
 $\ftwodfour$ in the process $ep \rightarrow eXp$   
in the range  $0.015 < \beta < 0.5$, 
$ 4\times 10^{-4}<\xpom<0.03$ and averaged over
$Q^2$ between 5 and 20 $\gevtwo$ and over $t$ in the interval $0.073<|t|<0.4~\gevtwo$. 
We also present a measurement of the differential cross
section as a function of $t$ in a similar  kinematic range.

\section{Experimental setup}

\subsection{HERA}

In 1994 HERA operated with 153 colliding bunches of $E_p=820$ GeV protons
and $E_e=27.5~\gev$ positrons. Additionally 15 unpaired positron  and 17
unpaired proton 
bunches circulated and were used to determine beam related background.
The integrated luminosity for the present study, which required the 
leading proton spectrometer to be in operating position (see below), 
is $900 \pm 14~\nbi$.

\subsection{The ZEUS detector}

A detailed description of the ZEUS detector is given elsewhere \cite{zeusdet}.
The main components of the  detector used in this analysis
are the uranium-scintillator calorimeter, the tracking detectors
and the leading proton spectrometer.

The uranium-scintillator calorimeter (CAL) 
covers $99.7\%$ of the solid angle and is
divided into three parts, the forward\footnote{The ZEUS 
coordinate system is right-handed with the $Z$-axis pointing in the
proton beam direction, referred to as forward, and the $X$-axis horizontal 
pointing towards the centre of HERA. The pseudorapidity $\eta$ is defined
as $-\ln(\tan \frac{\theta}{2})$, where the polar angle $\theta$ is measured with 
respect to the proton beam direction.}
(FCAL, covering the range
$4.3> \eta > 1.1$ in pseudorapidity), the barrel (BCAL, $1.1> \eta > -0.75$)
and the rear (RCAL, $-0.75> \eta > -3.8$). Each part is longitudinally 
subdivided into
electromagnetic (EMC) and hadronic (HAC) sections with typical cell sizes
of $5 \times 20 ~\cmtwo$ in the EMC ($10 \times 20 ~\cmtwo$ in the RCAL) and 
$20 \times 20 ~\cmtwo$ in the HAC.
The timing resolution is better than $1~{\rm ns}$ for energy deposits
greater than 4 GeV.
The energy resolution was measured in test beams \cite{calor} to be
$\sigma/E=18\%/{\sqrt E(\gev)}$ for electrons and
$\sigma/E=35\%/{\sqrt E(\gev)}$ for hadrons.
In order to minimize the effects of noise due to the uranium activity,
the isolated EMC (HAC) cells with energy less than 100 (150) MeV were discarded
from the analysis.

The tracking system consists of two concentric
cylindrical drift chambers, the  vertex detector (VXD) \cite{vxd} and the
central tracking detector (CTD) \cite{ctd}, operating in a magnetic 
field of 1.43~T. 
The CTD, which encloses the VXD, is a drift chamber consisting of 72
cylindrical layers, arranged in 9 superlayers.
The measured resolution in transverse momentum for tracks with hits in all
superlayers is $\sigma_{p_T}/{p_T} =0.005 p_T \bigoplus 0.016$
($p_T$ in GeV).
The interaction vertex is measured with a resolution
of 0.4 cm in $Z$ and 0.1 cm in the $XY$ plane.

The position of the scattered positron close to the rear beam pipe
region is determined with precision by the small-angle rear tracking detector
(SRTD), consisting of 2 planes of scintillator strips
 attached to the front face of the RCAL, covering an area of 
$68 \times 68 ~\cmtwo$. A hole of $20 \times 20 ~\cmtwo$ 
at the centre of the RCAL and the SRTD accommodates the beam pipe.
The SRTD signals resolve single minimum ionizing particles and provide
a position resolution of 0.3~cm. The time resolution is better than
2 ns for a minimum ionizing particle.
The SRTD is also used as a presampler to correct
the positron energy for losses in the inactive material in front
of the rear calorimeter \cite{f2svtx}. 

The proton remnant tagger (PRT1) \cite{zeusphp} is a set of scintillator counters
surrounding the beam pipe in the forward part of the ZEUS
detector at $Z=5~{\rm m}$. The tagger consists of two layers of
scintillating materials separated by a 1 mm thick lead absorber. 
Each layer is split vertically into two halves and each half is read out
by a photomultiplier tube. 
The geometric acceptance of PRT1 extends over the pseudorapidity
region $4.2 <\eta < 5.8$. 

The luminosity is determined via the bremsstrahlung process $e^+p \rightarrow
e^+p\gamma$  by measuring energetic photons in a lead-scintillator
calorimeter (LUMI) placed at $Z=-107 ~{\rm m}$ along the beam line \cite{lumi}.

The LPS  \cite{lpsrho} measures protons
scattered at very small angles with respect to the beam line
and escaping the central apparatus through the forward beam hole.
Such protons
carry a substantial fraction $x_L$ of the incoming proton 
momentum and have a small transverse momentum $p_T$ $(\sleq 1~\gev)$.
The spectrometer 
consists of six detector stations located at distances of $23$ to $90~{\rm m}$
along the proton beam line.
 In 1994 the three most forward stations  
S4, S5 and S6 were operational: 
each of these stations consists of an upper and a lower half,
which partially overlap during data-taking.
Each half 
is equipped with six rectangular parallel planes of
silicon micro-strip detectors.
Three different strip orientations 
(two vertical, two
at $+45^\circ$ with respect to the vertical direction 
and two at $-45^\circ$) are used,
in order to remove reconstruction ambiguities. 
The dimensions of the planes vary from station to station and are
approximately $4 \times 6 ~\cmtwo$; the pitch is $115~\mu{\rm m}$ for the
planes with vertical strips and $81 ~\mu{\rm m}$ for the other planes.
The edges of the detectors closest to the beam
have an elliptical contour which follows
the $10\sigma$ profile of the beam, where $\sigma$ is the standard deviation 
of the beam spatial distribution in the transverse plane.
During data-taking, the planes are inserted in the beam pipe 
by means of re-entrant Roman pots \cite{romanpot} 
and are retracted during beam dump and filling operations
of the HERA machine.
The LPS coordinates are reconstructed  with a precision of approximately
$35~ \mu {\rm m}$, which includes the intrinsic resolution of the coordinate
reconstruction and the alignment precision of the detector stations. 
The track deflection induced by the magnets in the proton
beam line is used for momentum analysis of the scattered proton.
The $\xl$ resolution was determined using elastic $\rho^0$ photoproduction 
events to be  better than $0.4\%$ at
$820~\gev$; the intrinsic $p_T$ resolution is approximately $5~\mev$
at $\xl=1$ and is less than the spread introduced by the 
angular divergence of the proton beam 
(which is $40~\mev$ in the horizontal and
$90~\mev$ in the vertical plane).

\section{Reconstruction of kinematic variables}

In order to reconstruct the kinematic variables $x,Q^2$ and $W$ (see section 2),
the so-called ``Double Angle'' method \cite{paulkoo} was used, which
derives the above quantities  
from the scattering angle of the positron and  that
of the struck (massless) quark. The latter  is deduced from the momenta of
all final state particles, except the scattered proton and positron. 


The final state particles in the reaction $e^+p \rightarrow e^+ X p$ were reconstructed
from the tracks and calorimeter energy deposits.  
The scattered positron identification algorithm was based on a neural
network \cite{sinistra94} which uses information from the CAL. 
The momenta of the particles of the system $X$
were reconstructed from calorimeter clusters and from
tracks in the CTD. 
The calorimeter clusters were formed by grouping CAL cells into 
cones around
the cell with a local energy maximum. The position of these objects was given
by the sum of the positions of the single cells 
weighted by the logarithm of their energy.
Reconstructed tracks were required to have transverse momenta 
of at least $100~\mev$ and were
matched to clusters with a procedure based on the distance of closest approach.
In case the cluster (track) was not matched to a track (cluster), the momentum of the particle 
was reconstructed only from the cluster (track). When the calorimeter cluster was matched
to a track, the track momentum measurement was taken if the following two
conditions were satisfied: the ratio between the energy of the cluster and the
track momentum was less than 0.8; the track momentum resolution
was better than the calorimeter energy resolution. If these two criteria were not
satisfied, the particle momentum was reconstructed from the calorimeter cluster energy.
An appropriate algorithm was developed for 
the cases in which more tracks
pointed to a single  cluster or more clusters were matched to one track.
The use of tracks improves the $M_X$ resolution and reduces the sensitivity to the 
losses due to inactive material in front of the calorimeter.

The quantity $t$ was determined 
from the fractional
momentum $x_L = p^{{LPS}}/E_p \simeq p_Z^{LPS}/E_p$ 
of the scattered proton measured in the LPS
and its transverse momentum $p_T^{LPS}$
with respect to the proton beam direction:
\begin{equation}
t =  -\frac{(p_T^{LPS})^2}{x_L}. 
\label{eqt}
\end{equation}
The $t$ resolution is approximately 
$\sigma(|t|)=140~\mev \sqrt{|t|}$
($t$ in GeV$^2$) and is dominated by the angular beam 
spread.

The quantities $\xpom$ and $\beta$ were reconstructed using 
the values of $Q^2$ and $W^2$ determined with
the double angle method and the mass of the final hadronic
system $M_X$:
\begin{equation}
\xpom = 
        \frac{M_X^2+Q^2}{W^2+Q^2} ,~~
\beta = 
 {\frac{Q^2}{Q^2+M_X^2}}.
\end{equation}

The mass $M_X$ was measured in two ways.
In the first method, it was reconstructed as:
\begin{equation}
(M_X^{meas})^2 = (\sum_h E^h)^2 - (\sum_h p_X^h)^2 - (\sum_h p_Y^h)^2 - (\sum_h p_Z^h)^2,
\end{equation}
where the sums run over all particles detected, except the scattered positron
and proton, and $\vec{p}^{~h}=(p_X^h,p_Y^h,p_Z^h)$ is the momentum vector assigned to each
particle of energy $E^h$.
From Monte Carlo studies the resolution on $M_X^{meas}$ 
was found to be approximately $40\%$ for $\mxz<3$~GeV, $18\%$
in the range between $3$ and $10$ GeV and $14\%$ for $\mxz$ between $10$ and $30$
GeV. The mean reconstructed $M_X^{meas}$ is $80\%$ of its true value for all values of 
$M_X^{meas}$:
therefore an overall correction factor of $1.25$ was applied. 
In the second method,
 $M_X$ was reconstructed using the LPS and the double angle variables:
\begin{equation}
\mxlpstwo = (1-\xl)(W^2+Q^2-m_p^2) -Q^2 + t,
\end{equation}
where $m_p$ is the proton mass.
When $\xl$ approaches unity (low $M_X$), the 
resolution on $\mxlps$ worsens. Monte Carlo studies showed a resolution
of $17\%$ for  $\mxlps>10$ GeV, of $35\%$ for $\mxlps$ between 3 and 10 GeV and
a substantially worse resolution for $\mxlps<3~\gev$. No
additional correction factor was needed.

In the analysis described here, the first reconstruction method
was used to evaluate the mass $M_X$, calculated 
as $1.25 \cdot M_X^{meas}$: the resulting resolutions on $\xpom$
and $\beta$ in the measured range are $30\%$ and $25\%$, respectively. 
A mixed method
with $\mxlps$ for high masses was used as a  systematic check.

\section{Monte Carlo simulation and LPS acceptance}

The diffractive single dissociation process,  $e^+p \rightarrow e^+ X p$,
 was modelled using two Monte Carlo (MC)
generators which assume pomeron exchange, the RAPGAP program and a program
based on the Nikolaev and Zakharov model, as well as a MC generator
for the exclusive reaction $e^+p \ra e^+ \rho^0 p$.

The RAPGAP \cite{rapgap} program is based on a factorizable model \cite{ingsch}
in which the incoming proton
emits a pomeron, whose constituents take part in the hard scattering.
For the pomeron flux, the  Streng parametrization \cite{streng} was used, 
which has an effective $t$ dependence ranging from $e^{-5|t|}$ to $e^{-9|t|}$
in the  $\xpom$ range covered by the present measurement; for the
pomeron structure function the form fitted in our previous analysis
\cite{zeusf2diff} was taken, which contains a mixture of a ``hard''
($\propto\beta(1-\beta)$) 
and a ``soft'' $(\propto{(1-\beta)}^2)$ quark parton density, with
no $Q^2$ evolution.
The program is interfaced to HERACLES \cite{heracles} 
for the QED radiative corrections, the parton shower is simulated
using the colour-dipole model
as implemented in ARIADNE \cite{ariadne} and
the fragmentation is carried out with the Lund string model as in JETSET
\cite{jetset}. The region of low masses was tuned to the measured ratio
of $\phi$ to $\rho^0$ resonance production \cite{zeusphi}.
The events were generated with a lower $\xl$ cut of 0.9 and with $|t|<1~\gevtwo$.

The Nikolaev and Zakharov model  
pictures the diffractive dissociation of the photon as a fluctuation of the photon
into a $q \bar q$ or $q \bar q g$ state \cite{nikzak}. The interaction
with the proton proceeds via the exchange 
of a two-gluon colour-singlet state. The  cross section can  be
approximated in terms of a two-component structure function of the pomeron,
each with its own flux factor.
The hard component, corresponding to the $q \bar q$ state,
has a $\beta$ dependence of the form $\beta(1-\beta)$ and an 
exponential $t$ distribution with a slope of approximately $10~\gevmtwo$. 
The soft component, which corresponds to the $q \bar q g$ state,
has a $\beta$ dependence of the form $(1-\beta)^2$ and
a $t$ slope  of about $6 ~\gevmtwo$.
In the Monte Carlo implementation of this model \cite{ada}, 
hereafter referred to as NZ, 
the mass spectrum contains both components but the
 $q \bar q g$ states are fragmented (using JETSET) 
into hadrons as if they were a 
$q \bar q$ system with the same mass. The generation is limited
to $M_X>1.7~\gev$. 

To improve the description of the low mass region, 
events in the region $M_X < 1.7~\gev$ were generated separately
using a MC for exclusive $\rho^0$ electroproduction, $e p \ra e \rho^0 p$.
In this MC the total cross section for
the process $\gamma^\ast p \ra \rho^0 p$ and the ratio of
the longitudinal to the transverse cross section are derived from a parametrization
of ZEUS \cite{rho93} and low energy data.
The generator 
is interfaced to HERACLES which simulates
 the initial and final state radiation.
The RAPGAP sample and the $\rho^0$ MC sample were mixed in such a way
that the weighted sum reproduced the observed $M_X$ distribution in the
data: only a small
contribution from the $\rho^0$ MC needed to be added
and this mixture was used for the acceptance correction. In a similar
way, the NZ and the $\rho^0$ MC were mixed and used 
as a systematic check.

In order to estimate the background,
the following processes were generated.
The events where the proton dissociates into a state of mass $M_N$
($ep \ra eXN$)
were generated with the EPSOFT~\cite{michal} and PYTHIA~\cite{pythia}
programs, where the mass spectrum of the nucleon system and the
ratio of double to single dissociation are
generated according to measurements from proton-proton colliders
\cite{goulianos}. The $M_N^2$ spectrum is of the form
$1/M_N^\alpha$, with $\alpha \simeq 2$.
To evaluate the background due to the one-pion
exchange process, the RAPGAP MC implementation of this exchange was used, 
in which the flux of pions is assumed to be
$f_\pi(x_\pi,t) \simeq (1-4 m_\pi^2)^2/(1-4t)^2 \cdot (x_\pi t / (t-m_\pi^2)^2)$
\cite{pion},
where $m_\pi$ denotes the pion mass and $x_\pi$ is the fraction of the proton
momentum carried by the pion.

All generated events were passed through the standard ZEUS detector
simulation, based on the GEANT program \cite{geant}, and through
the trigger simulation and the event reconstruction package. For the
scattered proton the simulation includes the geometry of the beam pipe apertures, 
the HERA magnets and their fields,  and the
response and noise of the LPS detectors. An accurate
simulation of the interaction vertex position and of the effect of the
proton beam tilt and emittance is also included.

The distribution of the detected proton transverse momentum, $p_T^{LPS}$, 
is limited by the 
geometric acceptance of the LPS, which depends on the
geometry of the beam pipe and the elliptical cutouts in the silicon
detector planes.
Figure 1 shows the LPS geometric acceptance for tracks with 
$\xl=1$ in the $p_X^{LPS},p_Y^{LPS}$ plane, where $p_X^{LPS},p_Y^{LPS}$ are the 
$X$ and $Y$ components of the proton momentum.
Protons with $p_T^{LPS} \sleq 0.2$~GeV and $\xl \simeq 1$ are too close to the beam to be 
safely measured.
 In the region covered by the LPS planes
($200 ~\mev \sleq       |p_X^{LPS}| \sleq 400 ~\mev, |p_Y^{LPS}| 
\sleq 600 ~\mev$)
the acceptance is large, as shown by the lines in the figure: however,
integrating over the azimuthal angle $\phi$
the resulting acceptance is $\simeq 6\%$.
For $\xl>0.97$ and $0.25 \sleq p_T^{LPS} \sleq 0.65$ GeV, the acceptance 
does not vary strongly with $x_L$ and
changes from $15\%$ at low $p_T^{LPS}$ to $2\%$ at high $p_T^{LPS}$. 
The alignment
uncertainty does not significantly affect the geometric acceptance, and, 
in this analysis, cuts were applied to ensure that the events lie in a well understood 
region.

\section{Data selection}

Deep inelastic events in the ZEUS detector were first selected 
online by a three-level trigger system 
(details can be found in \cite{f2nvtx}). 
At the first level DIS events were selected by requiring a 
minimum energy deposit in the electromagnetic section of the CAL. 
The threshold varied
between 3.4 and 4.8 GeV depending on the position in the CAL. 
At the second level trigger, beam-gas background
was reduced by using the measured times of energy deposits and the summed
energies from the calorimeter. Events were accepted if 
$\delta=\sum_i E_i(1-\cos\theta_i) > 24 ~\gev - 2 E_\gamma$,
where $E_i,\theta_i$ are the energies and polar angles of the calorimeter cells 
and $E_\gamma$ is the energy measured in the luminosity photon calorimeter, 
thereby accounting for the photon emitted in events with initial state 
radiation.
For fully contained events, the quantity $\delta$ is expected to be
twice the positron beam energy, $\delta \simeq 55 ~\gev$.
At the third level, algorithms to reject beam-halo 
and cosmic muons, as
well as a stricter $\delta$ cut ($\delta> 25 ~\gev-2 E_\gamma$), 
were applied,
together with the requirement of a scattered positron candidate with energy 
greater than $4~\gev$. 

Diffractive DIS candidate events were further selected offline in
two steps: first the standard inclusive
DIS selection was applied, then a high $x_L$ track in the LPS  was required.

Neutral current DIS events were selected as follows:
\begin{itemize}
\item{} A scattered positron with  
$E_e^\prime>10~\gev$  was required, 
where $E_e^\prime$ is the energy after presampler correction 
with the SRTD (when available). 
\item{} The impact point of the scattered positron in the SRTD was required
to be outside a square of $26 \times 26 ~\cmtwo$ centred on 
the beam. This cut ensures full containment of the positron shower in RCAL. 
\item{} $40 < \delta  < 65 ~\gev$, to reduce the photoproduction background
and the radiative corrections.
\item{} The $Z$ coordinate of the reconstructed 
vertex was required to be in the
range $-50 < Z < 100~{\mathrm cm}$.
\item{} $y_{e}<0.95$, where $y_{e}=1-E_e^\prime(1-\cos\theta_e)/2E_e$ is the 
value of $y$ 
calculated from the positron variables ($\theta_e$ is
 the polar angle of the scattered positron). 
\item{} $y_{JB}>0.03$, where $y_{JB}=\sum_h E_h(1-\cos\theta_h)/2 E_e$
is the value of $y$ calculated from the hadronic energy flow \cite{jacblo}:
in this case the combination of tracks and clusters 
described in section 4 was used.  
\end{itemize}

The following cuts were applied to select
diffractive events:
\begin{itemize}
\item{} A track in the LPS was required, with $0.97<x_L<1.01$, where the lower
limit was applied to reduce non-pomeron exchange contributions and to
select a region of uniform acceptance,
while the upper limit corresponds to a $+2.5 \sigma$ distance from the
$\xl=1$ peak, where $\sigma$ is the average LPS resolution at $\xl \simeq 1$.
\item{} 
The LPS track was extrapolated along the proton 
beam line. No track was accepted
if, at any point, the
minimum distance of approach 
to the beam pipe, $\Delta_{pipe}$, was less than $500~\mu {\rm m}$. This cut 
reduces the sensitivity
of the acceptance to the uncertainty in the position of the beam pipe apertures.
\item{} The total 
$E+ p_Z \simeq  (E+p_Z)_{\rm Cal}+2 p_Z^{LPS} =
\sum_i E_i (1+\cos\theta_i)+2 p_Z^{LPS}$ of the event 
(the sum of the energy and the longitudinal component of
the total momentum measured in the calorimeter and in the LPS) 
was required to be $<1655~\gev$.
For fully contained events this quantity should be equal to 
$2 E_p =1640~\gev$;
the cut, which takes into account the resolution on the
measured value of
$p_Z^{LPS}$, reduces the background due to overlay events
(see following section).
\end{itemize}

After this selection $553$ events were left of which 376 were in the
$Q^2$ region between $5$ and $20~\gevtwo$.

Figures 2 to 4 show the distributions of 
the selected events and of the Monte Carlo
model used in the analysis (weighted sum of the RAPGAP
and the $\rho^0$ MC samples), as a function of
$Q^2,x,W,y_{JB}$ (figure 2), of $\eta_{max},M_X,\xpom,\beta$ (figure 3) and of 
$\xl,|t|,p_X^{LPS},p_Y^{LPS}$ (figure 4).
The distributions are all uncorrected and the MC histograms 
are normalized to the number of events
in the data; a cut
of $Q^2> 5~{\mathrm GeV}^2$ was applied to both data and Monte Carlo samples.
In fig.~3a, the variable $\eta_{max}$ is the maximum pseudorapidity of
all calorimeter clusters with an energy of at least $400~\mev$ or tracks
with momentum of at least $400~\mev$. In our previous analysis,
 events with a large rapidity gap
were defined by $\eta_{max} \sleq 1.5$-$2.5$ \cite{zeusf2diff}:
however the events with a tagged proton with $\xl>0.97$ 
can have larger values of $\eta_{max}$. This allows us to study events with
$M_X \sgeq 20~\gev$, where the hadronic system extends
close to the proton beam pipe, and no gap may be observed in the detector.
In figs.~3c and~3d, the $\ln M_X^2$ distribution is shown in two $W$ bins
($50<W<120~\gev,120<W<270~\gev$). 
The data and the diffractive MC model are in 
good agreement.
Compared to our previous analyses, the
data extend the explored kinematic region to higher values of $M_X$, and
therefore to higher $\xpom$ and lower $\beta$.

\section{Background}

After the selection described in the previous section, 
the sample still contains some background, mainly due
to beam-halo and DIS processes which are not single diffractive
dissociation.

Proton beam-halo events
originate from interactions of beam protons with
the residual gas in the pipe or with the beam collimators.
These events have a scattered proton of energy close to that of the 
beam:  when they accidentally overlap with a genuine DIS event,
they may give a false diffractive signal.
Most of these events, however, appear to
violate energy and momentum conservation.
Figure 5 shows a scatter plot of
$\xl$ and the $(E+p_Z)_{\rm Cal}$ (measured with the calorimeter)
for the DIS events selected in the LPS (excluding the $x_L$
and the $E+p_Z$ cuts): a clear band at $\xl \simeq 1$, uncorrelated
with the energy measured in the calorimeter, can be ascribed to
beam-halo events. 
This type of background can be rejected by  requiring that the
total $E+p_Z$ of the event be conserved, as for beam-halo it
can exceed the kinematic limit of $1640~\gev$ (see line in fig.~5).
To evaluate the residual background after the cuts mentioned in the previous
section, the $2 p_Z^{LPS}$ distribution of events with unphysical tracks
in the LPS ($(E+p_Z)_{\rm Cal}+2 p_Z^{LPS}>1655~\gev$) was randomly mixed with the
$(E+p_Z)_{\rm Cal}$ distribution for DIS events, to create a $E+p_Z$
distribution for beam-halo events.
The obtained $E+p_Z$
distribution was normalized to the observed data distribution 
for $E+p_Z>1655~\gev$. 
The remaining background (below the $E+p_Z=1655~\gev$ cut) was estimated
to be less than $6\%$.

In order to evaluate the
background due to proton dissociation and 
pion exchange, the cut on $\xl$ was removed.
Figure 6 shows the observed uncorrected $\xl$ spectrum for the data (dots):
a narrow diffractive peak is seen at $x_L \simeq 1$, together with 
a distribution at lower $\xl$ 
due to the background processes mentioned above. Note that the acceptance
falls by almost an order of magnitude between $\xl \simeq 0.8$ and 
$\xl \simeq 1$.
These processes were modelled using the Monte Carlo programs
described in section 5, 
EPSOFT for the proton dissociation, 
RAPGAP for the pion exchange, while for the diffractive signal
the combined RAPGAP and $\rho^0$ MCs were used. 
The weighted sum of the three components was fitted to the observed $\xl$ spectrum
in the data in two steps.
The weight factor for the proton dissociation events was 
first determined by normalizing the EPSOFT MC sample  to the data
in the region $\xl<0.95$ and $\eta_{max}<1.5$, where double dissociation
dominates. The weight factors for 
the single diffraction  and the pion exchange processes were
determined by fitting the weighted sum of the three 
MC samples to the observed $\xl$ distribution.
The resulting sum is shown as the solid line in fig.~6.
After applying the $\xl>0.97$ cut, the background due to proton dissociation is less
than $3\%$, to
be compared with the 10-$15\%$ contamination
estimated to be present in our previous analyses \cite{zeusf2diff,kowalski}.
The fit was repeated using different MC models (PYTHIA for proton
dissociation, NZ plus $\rho^0$ for the signal) 
and consistent results were found.
This estimate of the background due to proton dissociation
was checked by looking at events tagged both
in the proton remnant tagger PRT1 and in the LPS. 
According to simulations based on the EPSOFT Monte Carlo, the PRT1 
has a tagging efficiency of approximately $50\%$ for proton dissociation events;
this leads, after the $\xl$ cut, 
to an estimated contamination below $3\%$,
consistent with the evaluation described above. 

The background due to the pion exchange was found to be less than 1\% 
after the $\xl>0.97$ cut.
Note that the fit to the $\xl$ spectrum is used to give
an estimate of the upper limit on the pion exchange
background at high $\xl$ and it is not 
meant to be a complete study of the distribution.
The background due to other reggeon exchanges was not included 
in this evaluation: at high $\xpom$ values ($\xpom \sgeq 0.01$),
a contribution from other reggeon exchanges is likely \cite{h1f2diff94}, 
however the
predictions vary significantly in different models \cite{golec,nikolaev}.
The possible contribution of these additional trajectories
at high $\xpom$ is discussed in section 9.
Background due to non-diffractive DIS processes (not shown in the figure), 
in which one of
the proton fragments is observed in the LPS, gives a contribution
at small values of $\xl$. 

The background due to non-$ep$ interactions (excluding beam-halo events)
was evaluated to be negligible from the data taken
with the unpaired proton bunches. The background due to
photoproduction events was found to be negligible from Monte Carlo
studies.

As the backgrounds due to beam-halo, non-$ep$ interactions, proton
dissociation and pion exchange were found to be small compared with 
the statistical precision of the data, 
they were not subtracted
in the results shown in the following.

\section{Measurement of the {\boldmath $t$} distribution} 

The measurement of the $t$ dependence in
diffractive DIS was limited to the kinematic range 
$0.073<|t|<0.4 ~\gevtwo$, $5 < Q^2< 20 ~\gevtwo$, $50 < W < 270~\gev$,
$0.015 < \beta < 0.5$ and $x_L>0.97$.
The range in $|t|$ was limited to values greater
than  $0.073~\gevtwo$, since  for
lower values of $|t|$  the acceptance varies rapidly;
the upper limit of  $0.4~\gevtwo$ 
restricts the data to a region where the LPS acceptance exceeds $2\%$.
The bin widths in $t$ were chosen to be larger than
the resolution,  resulting in four bins.
The acceptance and the
detector effects were unfolded  using a bin-by-bin correction determined
with the RAPGAP plus $\rho^0$ MC.

Figure 7 shows the measured differential cross section, $d\sigma/dt$.
The distribution was fitted with a single
exponential form, shown as the solid line: 
\begin{equation}
d\sigma/dt = A e^{bt};   
\end{equation}
the value of the fitted parameter $b$ is:
\begin{equation}
b= \tslopeerr,  
\end{equation}

where the first uncertainty is statistical and the second is systematic.

The systematic uncertainties can be subdivided into three groups,
those due to the DIS selection, those due to the LPS acceptance
and background and, finally, those related to the unfolding.
The systematic error due to the DIS acceptance
was evaluated by changing the following cuts:
the positron energy cut was
moved to $8~\gev$ and $12~\gev$, 
the cut on the impact position of the positron using the SRTD
was moved to $24 \times 24 ~{\rm cm^2}$ and to $28 \times 28 ~{\rm cm^2}$; 
the $y_{JB}$ cut was changed to 0.02 and 0.04. The effects on the
slope $b$ varied between $+5.5\%$ and $-3\%$.
To estimate the systematic contributions due to the LPS acceptance,
the following checks were performed:
the $\Delta_{pipe}$ cut was increased to $0.1~{\rm cm}$ causing a
negligible variation to the slope. The values of $p_X^{LPS}$ and $p_Y^{LPS}$
were shifted by $\pm 3~\mev$ and $\pm 6~\mev$, respectively, as a test
of the influence of the alignment procedure  
(see~\cite{lpsrho}), yielding
variations on $b$ between $-6\%$ and $+3\%$.
The mass $M_X$ was reconstructed
using the calorimeter for low masses and the LPS for high masses,
giving changes around $\pm 5\%$ on the slope.
The NZ plus $\rho^0$ 
Monte Carlo was used for the acceptance calculation, the $t$ slope
in RAPGAP was changed by $\pm 1~\gevmtwo$, the $\aprime$ value in the MC
was changed from $0.25~\gevmtwo$ to $0$. 
These changes lead to variations of at most $4\%$ for the slope $b$. 
The total systematic error was obtained by adding in quadrature the
positive and negative deviations separately. 

In a Regge-type approach \cite{mueller}, the 
slope of the exponential $t$ distribution
in single diffractive interactions is predicted to
be  
$b \simeq b_0 + 2 \aprime \ln (1/\xpom )$. 
Assuming the value of $b_0=4.6~\gevmtwo$ inferred by Goulianos \cite{goulianos2}
from elastic $\bar p p$ scattering
at $\sqrt s =1800~\gev$ \cite{cdf2}
and  $\aprime=0.25~\gevmtwo$ \cite{dl}, 
in the present kinematic range ($<\xpom> \simeq 8 \times 10^{-3}$) 
the measured $b$ value is compatible 
with the predicted value of $ \simeq 7~\gevmtwo$ for a soft pomeron exchange;
it is also consistent 
with the values predicted by some perturbative QCD models
\cite{nikzakzol}.

\section{Measurement of {\boldmath $\ftwodfour$} and {\boldmath $\ftwod$}}

The measurement of the proton in the LPS 
permits the determination of the cross section for single diffractive dissociation 
in DIS as a function of the four kinematic variables 
$\beta,Q^2,\xpom ~{\rm and}~ t$. Given the small statistics
collected in the 1994 running, the measurement was performed 
in a single $t$ bin, $0.073< |t| < \ftwodfourtmax ~\gevtwo$ and a single
$Q^2$ bin, $5 < Q^2< 20~\gevtwo$, with average values of 
$<|t|>=0.17~\gevtwo$ and $<Q^2>=8~\gevtwo$.
The sizes of the $\beta$ and $\xpom$ bins  
were chosen to be larger than the resolutions (see section 4);
a minimum number of 8 events in each bin was also required.
The chosen bins are shown in figure~8:
they are in the range 
$4 \times 10^{-4} < \xpom < 0.03, 
0.015 < \beta <0.5$.
The acceptance, which includes the efficiency of the DIS
selection and the geometric acceptance of the LPS, varies between
$3\%$ and $11\%$ in these bins.
The purity, defined as the ratio of the number of
MC events generated in a bin and 
reconstructed in the same bin over the 
number of events reconstructed 
in that bin,
was required to be greater than $25\%$ in the bins used for the
measurement and is typically
more than $40\%$.
The effect of the longitudinal structure function $F_L$ is 
assumed to be smaller than the statistical errors
in the kinematic range considered \cite{zeusf2diff} and is 
neglected.

In order to extract the structure function $\ftwodfour$,
the weighted sum of the RAPGAP and the $\rho^0$ MC, 
which gives a good description of the data, was used to
obtain the correction for acceptance with a bin-by-bin unfolding method.
In addition a bin centring
correction was applied, using the
$\ftwodfour$ parametrization of the RAPGAP MC.

The following systematic checks were performed for the measurement 
of $\ftwodfour$:

\begin{itemize}
\item{} To check the photoproduction background and the QED 
        radiative corrections,
        the cut on $\delta$ was changed to $37~\gev$ and $42~\gev$,
        and the cut on the scattered
        positron energy was varied to $8~\gev$ and $12~\gev$. 
        The effect was typically up to $3\%$.
\item{} To check the acceptance at low $Q^2$ which is determined
         by the positron position, the cut on the impact point of the positron 
         was changed to $24 \times 24~{\rm cm^2}$ and to $28 \times 28~{\rm cm^2}$;         
         both induced variations on $\ftwodfour$ smaller than $15\%$.  
\item{} An alternative method to reconstruct the kinematics, the
        $\Sigma$ method \cite{h1sigma}, was used, giving typical variations
        of about $20\%$.
\item{} The cut on $y_{JB}$ was moved by $\pm$ 0.01, yielding changes 
        of typically around $10\%$.
\item{} The cut on the minimum energy deposit of the EMC (HAC) cells was moved to
         140 (160) MeV and to 80 (120) MeV, to check  the effect of
         the calorimeter noise on the mass reconstruction. The effect was between
         $10\%$ and $25\%$ for the first cut, and up to $15\%$ for the second check.
\item{} The LPS track selection was modified by 
        raising the $\Delta_{pipe}$ cut to $0.1~\cm$, 
        producing variations of $\ftwodfour$ of less than $12\%$.
\item{} To check the background estimation due to
        one-pion exchange and proton dissociation,
        MC predictions were statistically subtracted from the bins, producing
        changes up to $5\%$ in a few bins.  The beam-halo was also
        subtracted         with negligible effects on $\ftwodfour$. 
        The $\xl$ cut was moved to 0.96, causing changes on
        $\ftwodfour$ of less than $5\%$, except in one bin at high
        $\xpom$ where the change was $15\%$.
\item{} An alternative method to reconstruct the mass $M_X$ was employed,
        which uses $1.25 \cdot M_X^{meas}$ at low masses and $\mxlps$ at high masses, giving
        an effect of less than $2\%$ at high $\beta$ and up to $25\%$ at 
        low $\beta$.
\item{} 
  An unfolding method based on Bayes' theorem \cite{giulio} was applied 
        for the acceptance correction causing variations typically less than $3\%$.
        The NZ plus $\rho^0$ Monte Carlo was used to unfold, giving changes
        typically less than $20\%$, except at low $\beta$ where
        the changes were         up to  $40\%$.  
        The $\xpom$ and $t$ dependence in RAPGAP were reweighted to 
        a fixed $t$ dependence of the form $(1/\xpom)
        e^{-7|t|}$ or to a fixed $\xpom$ dependence of the form 
        $(1/\xpom)^{1.3}$, yielding changes of less than $5\%$.

\end{itemize} 
Most of the systematic checks yielded results which agree with the 
central value within the statistical errors.
The negative and positive deviations in each bin were separately
combined in quadrature.

The results for $\ftwodfour$ are given in table 1. The systematic errors do not
include a $5.5\%$ overall normalization uncertainty due to the luminosity
determination and trigger efficiency ($2\%$) and due to the 
LPS acceptance varying because, due to the proton beam conditions,
the LPS stations had to be positioned differently in different runs ($5\%$).
The values of $\ftwodfour$ are plotted
in fig.~9 as $\xpom \cdot \ftwodfour$ 
as a function of $\xpom$, in four $\beta$ intervals, with central values of 
$\beta=0.028,0.07,0.175$ and $0.375$, respectively. 
The $\ftwodfour$ data are observed to fall rapidly with increasing $\xpom$.
To investigate whether the $\xpom$ dependence is the same in all $\beta$
intervals, we performed fits  
of the form:
\begin{equation}
\ftwodfour = A_i \cdot \xpoma,
\label{xpomfit}
\end{equation}
where the normalization constants $A_i$ were allowed 
to vary in each bin, while the 
exponent $a$ was constrained to be the same in all four $\beta$ bins.
The result of the fit is 
\begin{equation}
a(0.073<|t|<0.4~\gevtwo) = \ftwodfouraerr,
\end{equation}
with a $\chi^2$ value of 10 for 8 degrees of freedom, showing that the
result is consistent with a single $\xpom$ dependence in all $\beta$ bins.
The systematic error was obtained by re-fitting the $\ftwodfour$ values
obtained for each of the systematic checks mentioned above. 
The main contributions arose from
changing the unfolding procedure (4\%),
using the $\Sigma$ method (5\%)
and applying different noise suppression cuts (7\%).

In order to compare with previous measurements of the diffractive
structure function, we extracted
$\ftwod(\beta,Q^2,\xpom)$ by integrating 
$\ftwodfour$ over $t$. 
The extrapolation to the whole $t$ range ($0 < |t| < \infty$) 
was performed using the $\ftwodfour$ parametrization in the MC. 
The result is shown in table 1 and  is plotted
as $\xpom \cdot \ftwod$ in figure 10. Fitting our data in each $\beta$ interval with
the form $(1/\xpom)^{\overline a}$ yielded a value $\overline a = \ftwodthreeaerr$.

This value of $\overline a $ is much lower than the value obtained 
with the ZEUS 1993 data \cite{zeusf2diff},  based on a large rapidity gap
analysis\footnote{We do not compare here to the result in ref.~\cite{kowalski},
as this result was affected by a technical error in the
Monte Carlo generation used for the unfolding, which led to a
higher $\bar a$ value by about one unit of the quoted error.}.
 The ZEUS 1993 $\ftwod$ result corresponds to 
$0.1 < \beta <0.8$  and
$6.3 \times 10^{-4} < \xpom < 10^{-2}$, for $8<Q^2<100~\gevtwo$,
and it is
compatible with a single $\xpom$ dependence in all
$\beta$ bins, with slope 
$\overline a=1.30~\pm 0.08 \mbox{(stat.)}^{+0.08}_{-0.14} \mbox{(syst.)}$.
As already mentioned, the present analysis covers a different kinematic
range,
extending to lower $\beta$ and higher $\xpom$ at $<Q^2>=8~\gevtwo$.
The lower value of $\bar a$ compared to our previous result 
may be ascribed to the presence of additional subleading trajectories contributing
in the $\xpom$ range covered by this analysis. 
This contribution has been already observed 
in the analysis of the H1 1994 data \cite{h1f2diff94},
which are shown for comparison in fig.~10. Only the points at the $Q^2$ 
and $\beta$ values
closest to the ones presented here are shown.
Assuming pomeron exchange to be the dominant 
contribution,
the values of $a(0.073<|t|<0.4~\gevtwo)$ and $\overline a$ can be related,
respectively, 
to the $\apom$ value in the given $t$ range and $\overline{\apom}$, 
the pomeron trajectory averaged over $t$ (see section 2).
However, as already mentioned in section 7, at high $\xpom$  
a contribution from reggeon exchange cannot be excluded, 
in addition to pomeron exchange:
the statistical precision of our data does
not allow us to identify this component.
As the reggeon trajectory has an intercept $\alpha_M \simeq 0.5$, leading
to an $\xpom$ slope $a \simeq 0$, its 
presence would lower the expected $\xpom$ slope 
 in the $\xpom$ region where
the reggeon exchange contributes. 
The $\xpom$ range covered by this analysis could then explain
the different result on the slope $\bar a $ compared to our previous result.
Moreover, models where the pomeron component is not factorizable
can lead to a different $\xpom$ slope parameter, depending on the 
kinematic range.

We have also examined the $\beta$ dependence of the diffractive structure function.
In order to do this, the $\ftwod$ values within each 
$\beta$ bin are plotted in figure~11, extrapolated to a common value of
$\xpom=0.01$, assuming a universal, 
fixed slope 
$\overline a=1.01$.  
The solid line represents a fit of the form
$b \cdot (1/\xpom)^{\bar a} [\beta(1-\beta) + c/2 (1-\beta)^2]$ to the measured
$\ftwod$. The values obtained in the fit are: 
$b=0.087 \pm 0.015~\mbox{(stat.)}^{+0.014}_{-0.022} \mbox{(syst.)}$ and
$c=0.34~\pm 0.11~\mbox{(stat.)}^{+0.25}_{-0.10} \mbox{(syst.)}$.
This fit to the present data indicates that both a hard ($\propto \beta(1-\beta)$)
and a soft ($\propto (1-\beta)^2$) component
are needed.
The parametrization 
obtained in our previous analysis \cite{zeusf2diff},  extrapolated to 
$\xpom=0.01$
and scaled down
by $15\%$ to take into account the proton dissociation background
contribution,  
is also shown in the figure as the dashed curve.

\section{Conclusions}

Diffractive DIS events have been studied at HERA using the ZEUS leading
proton spectrometer. A clean sample of events was selected by requiring
a scattered proton with $\xl>0.97$. 
The background due to proton dissociation was estimated to be
approximately $3\%$, substantially lower than in our previous analyses.
The use of the LPS has also allowed an extension of the ZEUS measurements to 
values of the final state hadronic mass $M_X$ as high as 35 GeV.

The $t$ dependence was measured for
the first time in this process in the range $0.073<|t|<0.4~\gevtwo$,
$5<Q^2<20~\gevtwo$, $0.015<\beta<0.5$ and $50<W<270~\gev$. 
The resulting distribution
is described by the function $e^{bt}$,
with $b=\tslopeerr$. 

The diffractive structure function $\ftwodfour(\beta,Q^2,\xpom,t)$
was measured in the interval $0.015<\beta<0.5$,
$ 4 \times 10^{-4} <\xpom < 3 \times 10^{-2}$ and averaged
over the range $0.073< |t| <0.4~\gevtwo$ and $ 5 < Q^2 < 20 ~\gevtwo$.
Because of the limited statistical precision of the data,
it is not possible to determine whether
a different $\xpom$ dependence is
needed in different $\xpom$ and $\beta$ ranges. 
The $\xpom$ dependence is consistent, in all
$\beta$ intervals,
with the form  $(1/\xpom)^{a}$, with
$a(0.073<|t|<0.4~\gevtwo)=  \ftwodfouraerr$. 
Integrating over $t$, 
the structure function $\ftwod$ was determined.
A fit of the form $(1/\xpom)^{\bar a}$ 
to $\ftwod$ yielded
$\overline a=\ftwodthreeaerr$.

The result for the effective $\xpom$ slope is lower than that obtained
in our previous measurement.
This analysis, however, extends the $\ftwod$ measurement to values
of $\xpom$ up to $0.03$, where
a significant component of reggeon exchange could contribute to
lowering the effective $\xpom$ slope parameter $\overline a$.

%
\section*{Acknowledgements}
We thank the DESY Directorate for their strong support and
encouragement. We acknowledge the assistance of the DESY
computing and networking staff.  
We are very grateful to the HERA machine group: collaboration with them
was crucial to the successful installation and operation of the LPS.
We would like to thank B.~Hubbard for his
invaluable contributions to the experiment, and the LPS in particular.
It is also a pleasure to thank N.N. Nikolaev and M.G. Ryskin 
for useful discussions.


\newpage

\begin{table}[htb]
\vspace{-0.cm}
\begin{center}
\begin{tabular}{|l|c|c|c|c|c|l|}  \hline
\bf{$\beta$}&\bf{$x_{I\!P}$}&\bf{$N_{obs}$}&\bf{$F^{D(4)}_{2}\pm stat.\pm sys.~\gevmtwo$}&
                                                    $ F^{D(3)}_{2} \pm stat. \pm sys.$\\  \hline\hline
0.028 & 0.011  & 14 &  $2.8 \pm 0.8  ^{+ 1.2}_{- 0.6}$ &  $1.3  \pm 0.4  ^{+0.6}_{-0.3}$ \\
0.028 & 0.024  & 17 &  $2.2 \pm 0.6  ^{+ 1.0}_{- 0.9}$ &  $1.0  \pm 0.3  ^{+0.5}_{-0.4}$ \\
0.07  & 0.0044 & 13 &  $7.8 \pm 2.3  ^{+ 2.0}_{- 2.1}$ &  $3.6  \pm 1.1  ^{+1.0}_{-1.0}$ \\
0.07  & 0.011  & 22 &  $3.8 \pm 0.9  ^{+ 1.3}_{- 1.4}$ &  $1.8  \pm 0.4  ^{+0.6}_{-0.7}$ \\
0.07  & 0.024  & 13 &  $1.8 \pm 0.5  ^{+ 0.7}_{- 0.3}$ &  $0.8  \pm 0.2  ^{+0.4}_{-0.1}$ \\
0.175 & 0.0018 & 15 & $28.4 \pm 7.7  ^{+13.7}_{- 4.8}$ & $13.4  \pm 3.6  ^{+6.5}_{-2.1}$ \\
0.175 & 0.0044 & 21 & $10.8 \pm 2.5  ^{+ 3.4}_{- 1.7}$ &  $5.0  \pm 1.2  ^{+1.8}_{-0.9}$ \\
0.175 & 0.011  & 22 & $5.3  \pm 1.2  ^{+ 1.2}_{-1.4}$  &  $2.5  \pm 0.6  ^{+0.5}_{-0.6}$ \\
0.175 & 0.024  & 13 & $3.6  \pm 1.1  ^{+ 1.0}_{- 1.7}$ &  $1.7  \pm 0.5  ^{+0.7}_{-0.7}$  \\
0.375 & 0.0007 & 19 & $89.1 \pm 21.7 ^{+14.3}_{-18.6}$ & $42.7 \pm 10.4  ^{+6.5}_{-6.4}$ \\
0.375 & 0.0018 & 32 & $53.6 \pm 10.4 ^{+8.6}_{-11.2}$  & $25.3  \pm 4.9  ^{+4.3}_{-4.6}$\\
0.375 & 0.0044 & 16 & $10.4 \pm 2.7  ^{+1.8}_{-4.2}$   &  $4.9  \pm 1.3  ^{+0.9}_{-1.8}$ \\
0.375 & 0.011  & 8  & $3.9  \pm 1.4  ^{+2.2 }_{-1.0}$  &  $1.8  \pm 0.7  ^{+1.0}_{-0.3}$ \\ \hline
\end{tabular}
\end{center}
\caption{Results on $\ftwodfour$ and $\ftwod$ using the ZEUS LPS 1994 data at $Q^2=8~\gevtwo$. 
The $\ftwodfour$ values are in the $t$ interval $0.073<|t|< 0.4~\gevtwo$, the $\ftwod$
values are in the $t$ range $ 0 <|t| < \infty$.
The overall normalization uncertainty of $5.5\%$ is not included and no
background is subtracted from the data.
}
\end{table}


\clearpage
\begin{figure}
\begin{center}
\leavevmode
\hbox{%
\epsfysize = 6.in
\epsffile{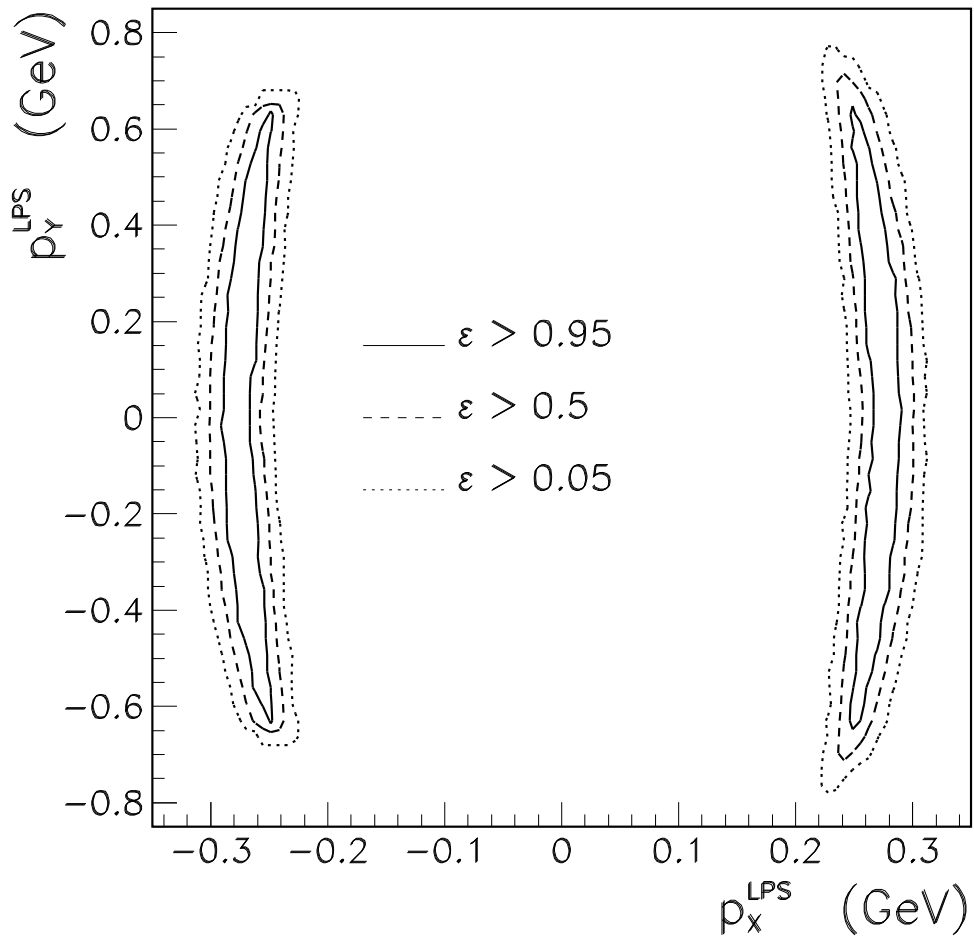}}
\end{center}
\caption{Geometric LPS acceptance for tracks with $\xl=1$ 
in the $p_X^{LPS},p_Y^{LPS}$ plane. 
The lines indicate the regions where the acceptance is 
greater than $5\%,50\%$ and $95\%$.} 
\end{figure}

\clearpage
\begin{figure}
\centering
\leavevmode
\hbox{%
\epsfxsize = 7.in
\epsffile{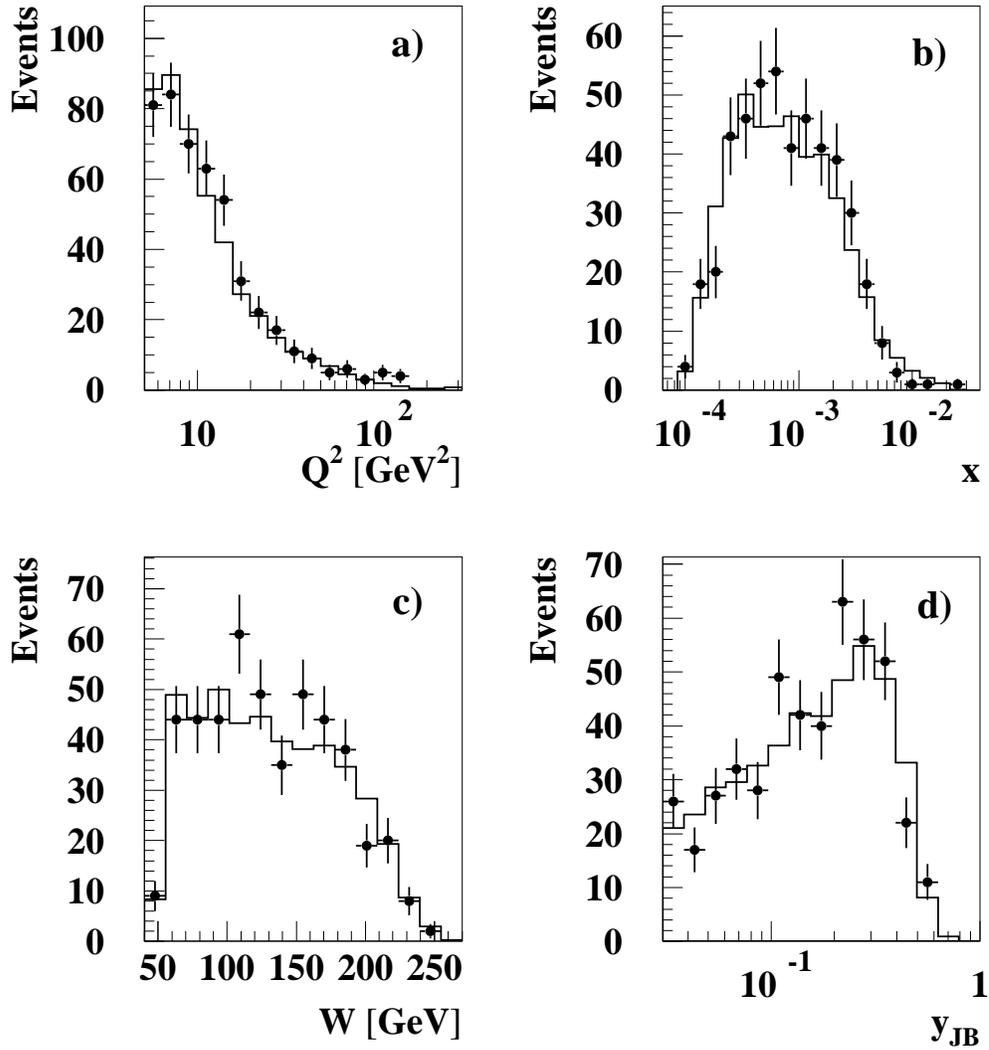}}
\caption{Observed event distributions as a function of (a) $Q^2$, (b) $x$, (c) $W$,
and (d) $y_{JB}$  of
the reconstructed data (dots) compared to the Monte Carlo model  
(solid line). 
The Monte Carlo is the weighted sum of the RAPGAP and $\rho^0$ samples. 
The errors, shown as vertical bars,  are statistical only.
}
\end{figure}

\clearpage
\begin{figure}
\leavevmode
\begin{center}
\hbox{
\epsfxsize = 6.in
\epsffile{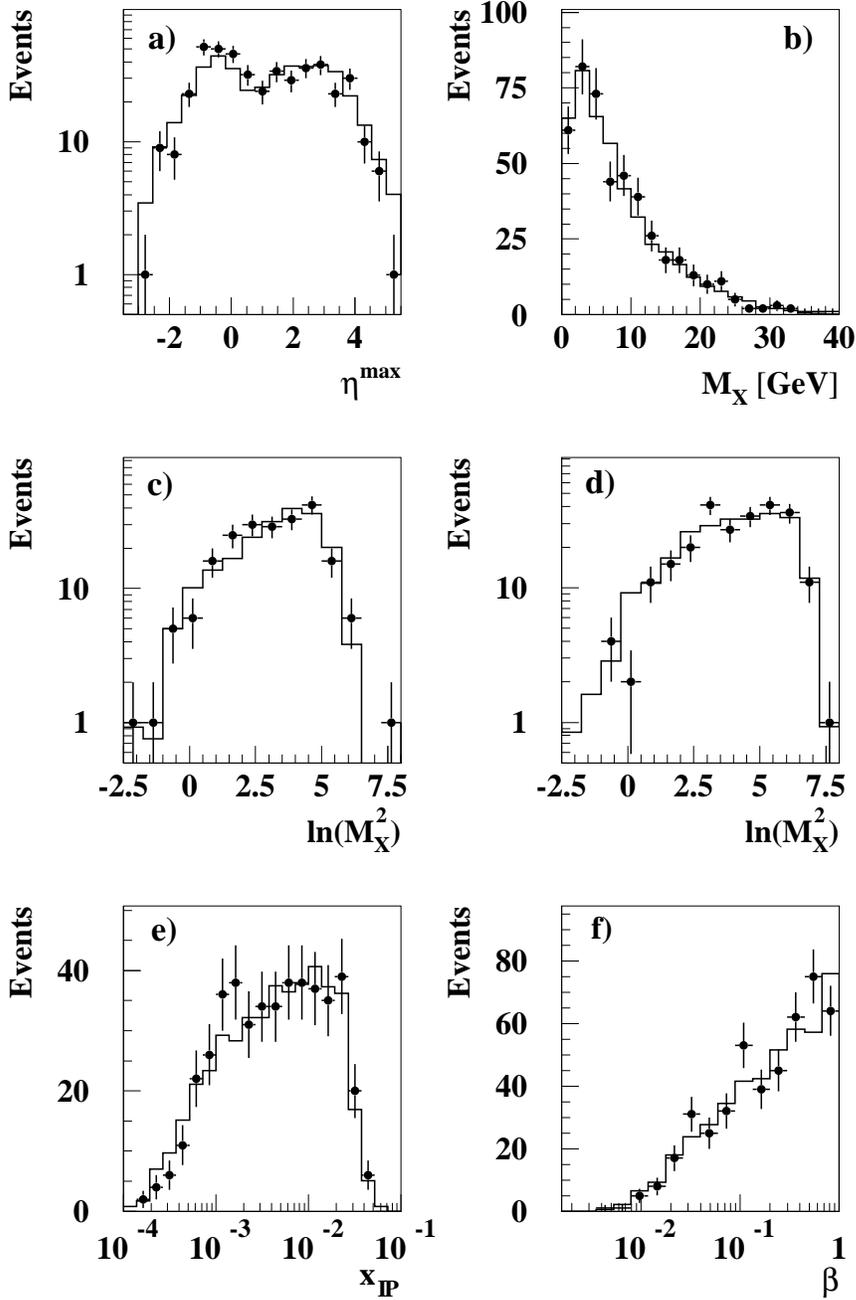}}
\end{center}
\vspace {-2cm}
\caption{Observed event distributions as a function of  (a) $\eta_{max}$  , (b) $M_X$,
(c) $\ln M_X^2$ for $50< W< 120~\gev $, (d) $\ln M_X^2$ for $120 < W <270 ~\gev $,
(e) $\xpom$ and (f) $\beta$ of
the reconstructed data (dots) compared to the Monte Carlo 
(solid line). 
The Monte Carlo is the weighted sum of the RAPGAP and $\rho^0$ samples. The errors are statistical only.
}
\end{figure}

\clearpage
\begin{figure}
\leavevmode
\begin{center}
\hbox{
\epsfxsize = 7in
\epsffile{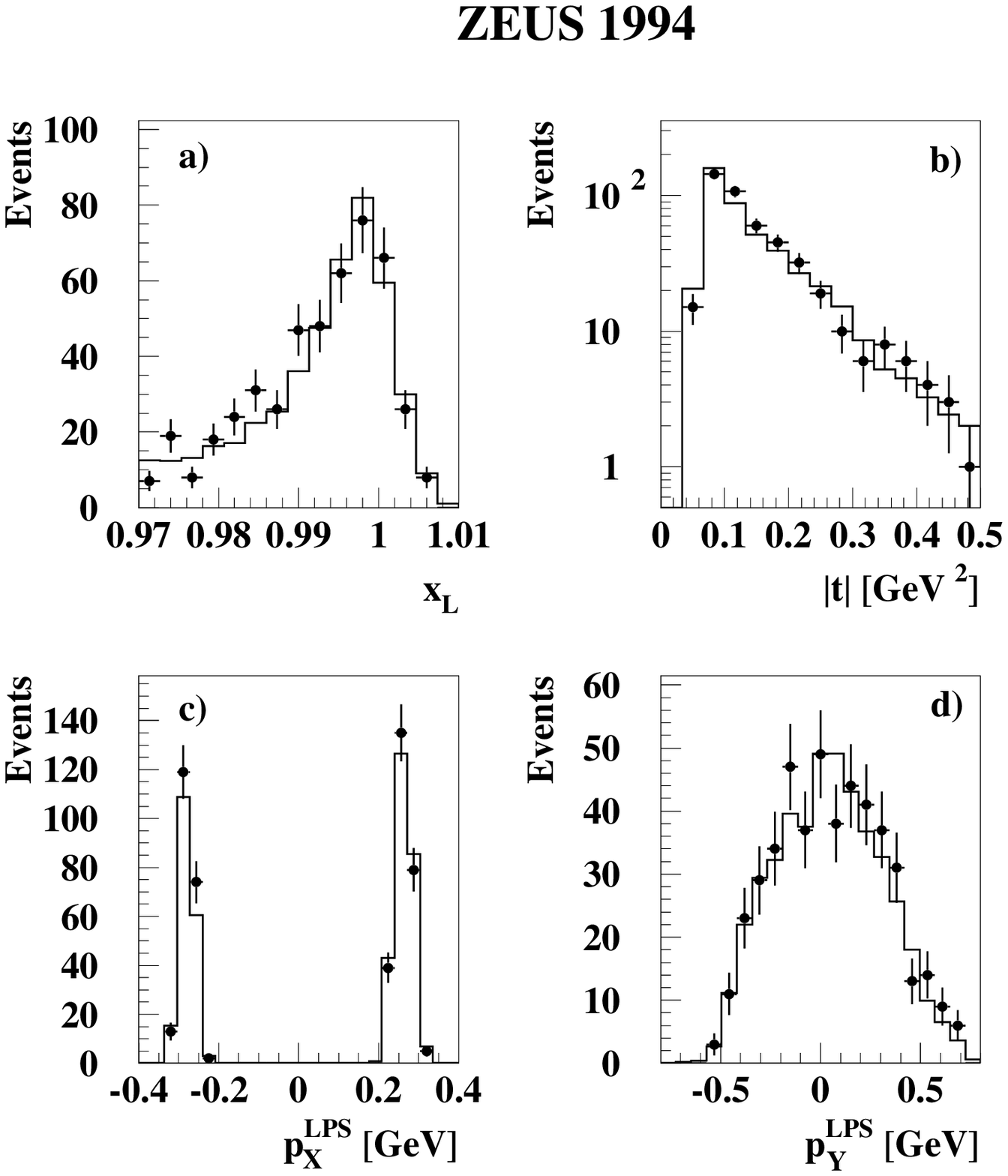}}
\end{center}
\vspace {-2cm}
\caption{Observed event distributions as a function of (a) $\xl$  , (b) $|t|$,
(c) $p_X^{LPS}$ and (d) $p_Y^{LPS}$ 
 of
the reconstructed data (dots) compared to the Monte Carlo
(solid line). 
The Monte Carlo is the weighted sum of the RAPGAP and $\rho^0$ samples.
The errors are statistical only.
}
\end{figure}

\clearpage
\begin{figure}
\begin{center}
\leavevmode
\vspace{1cm}
\hbox{
\epsfxsize = 5.5in
\epsffile{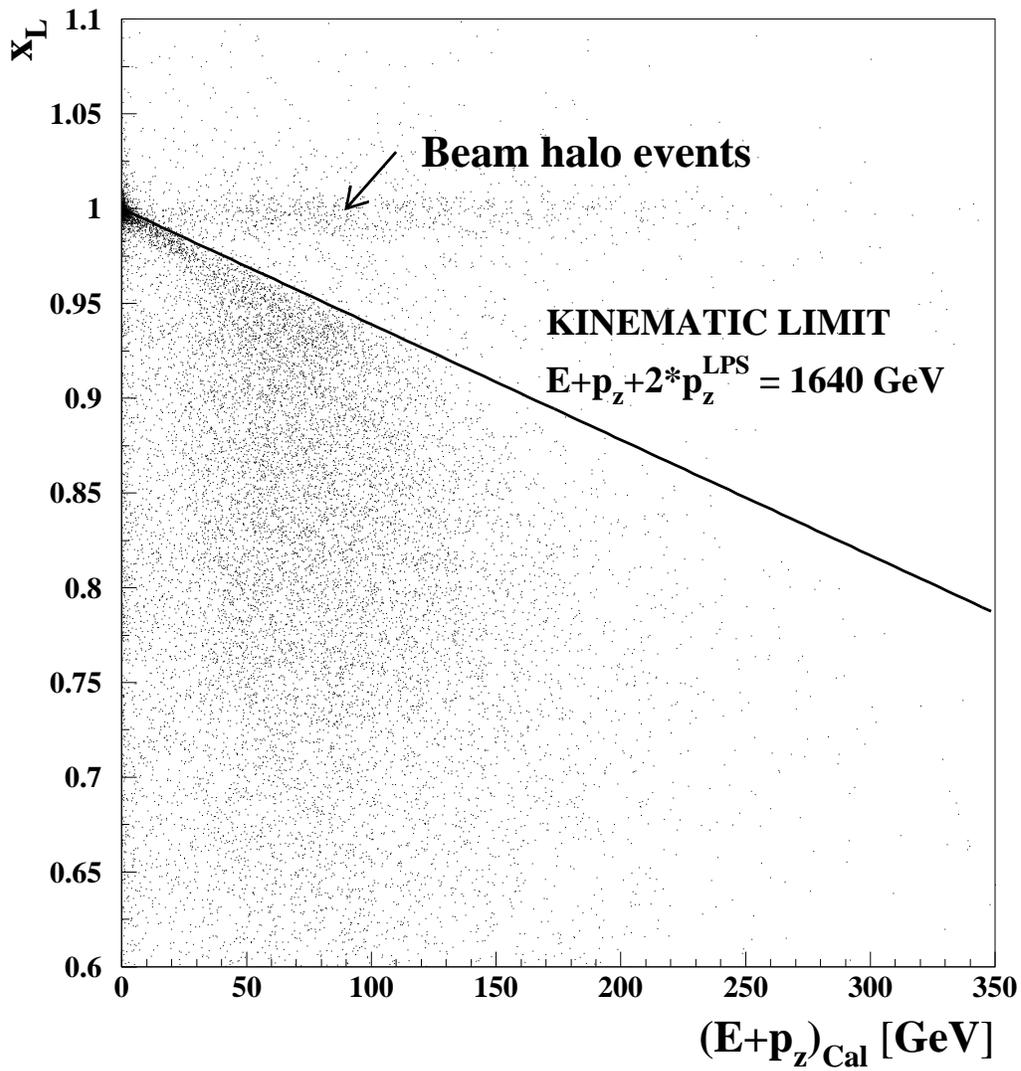}}
\end{center}
\vspace{-2cm}
\caption{
A scatter plot of $\xl$ and the $(E+p_Z)_{\rm Cal}$ as 
measured in the calorimeter in the data (before the $\xl$ and the $E+P_Z$ cut).
The region below the solid line indicates the allowed kinematic region.
} 
\end{figure}

\clearpage
\begin{figure}
\begin{center}
\leavevmode
\hbox{%
\epsfysize = 5in
\epsffile{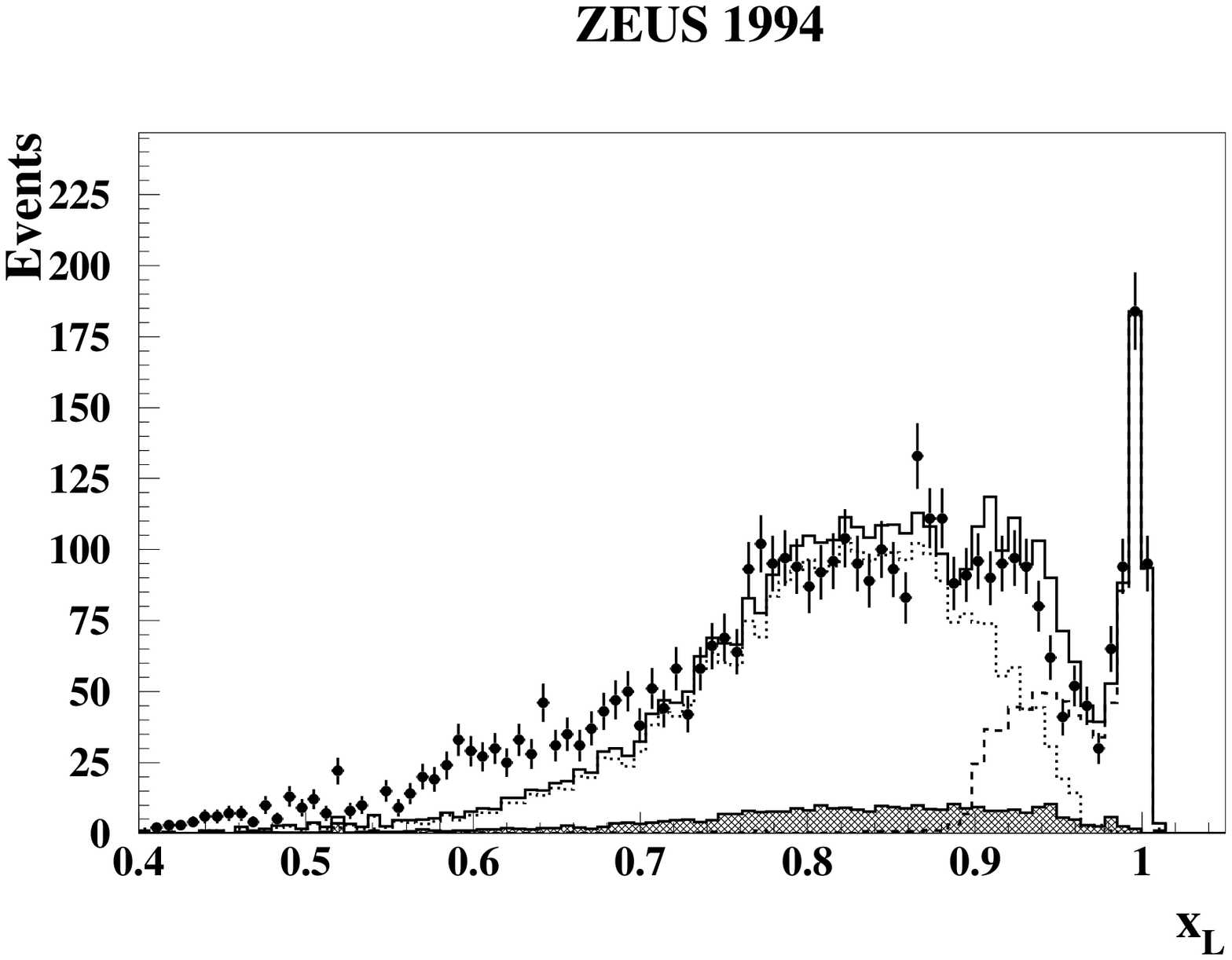}}
\end{center}
\vspace{1cm}
\caption{ The observed $\xl$ spectrum 
in the data (dots) where the $\xl$ cut at 0.97 has been removed.
Overlaid is the result of fitting the distribution with a sum (full line)
of the contribution due to proton dissociation (EPSOFT MC -- shaded area),
of the maximum contribution due to pion exchange (dotted line)
and of the single photon dissociation signal
(RAPGAP plus $\rho^0$ MC -- dashed line).
} 
\end{figure}

\clearpage
\begin{figure}
\begin{center}
\hbox{%
\hspace{2cm}
\epsfysize= 18cm
\epsffile{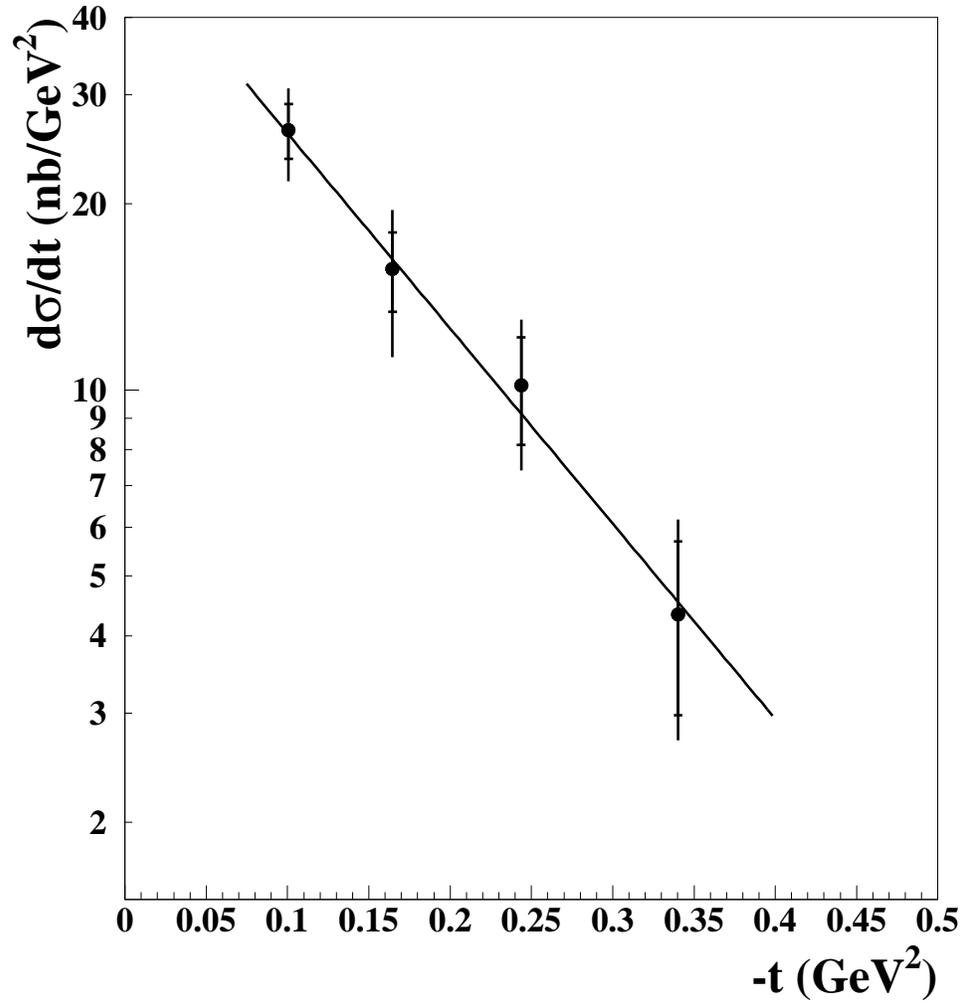}}
\end{center}
\caption{ 
The differential cross section $d\sigma/dt$ for diffractive DIS events with
a leading proton of $\xl>0.97$,
in the range $5<Q^2<20~\gevtwo$, $50<W<270~\gev$ and $0.015<\beta<0.5$.
The inner error bars indicate the statistical errors, the outer
error bars show the statistical and systematic errors added in quadrature.
The line is the
result of the fit described in the text.
}
\end{figure}

\clearpage
\begin{figure}
\begin{center}
\leavevmode
\hbox{%
\epsfxsize = 6.5in
\epsffile{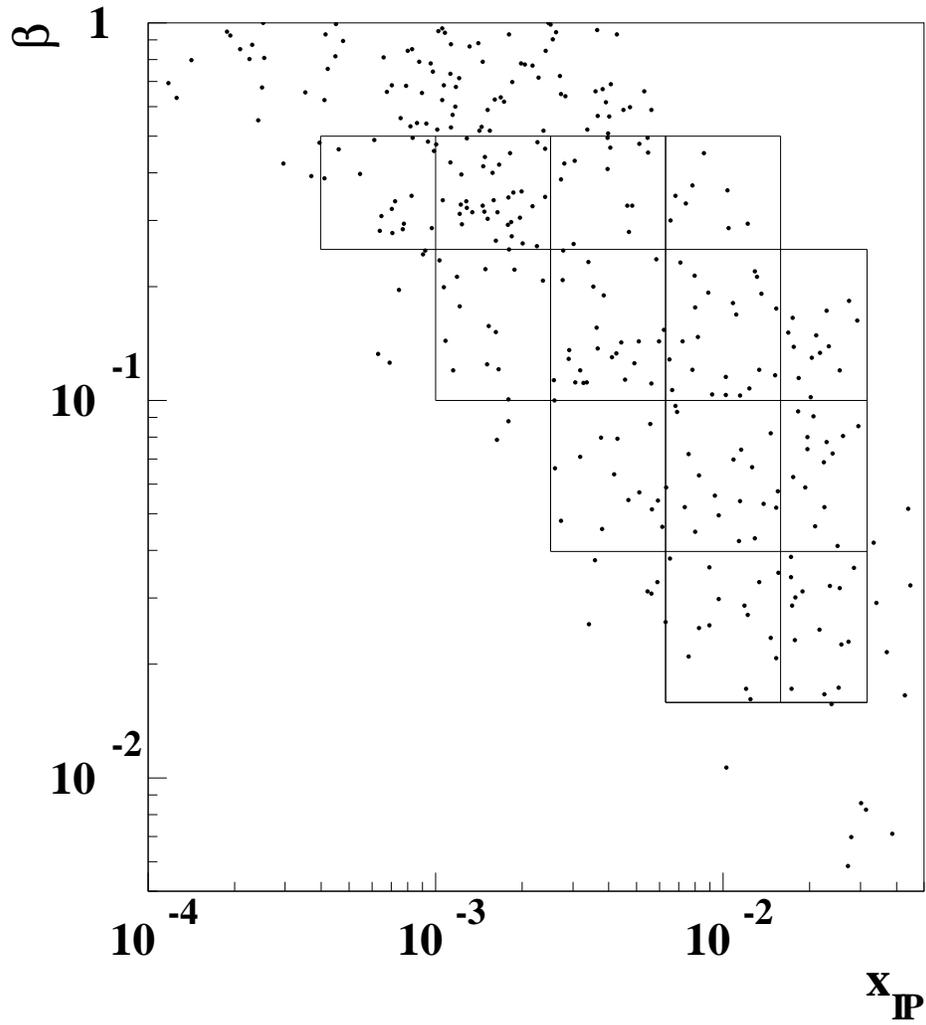}}
\end{center}
\caption{ 
The distribution of the events used in the analysis in the $\beta,\xpom$
plane. 
The bins in $\beta,\xpom$ used for the extraction of $\ftwodfour$ and $\ftwod$ are
shown.
}
\end{figure}

\clearpage
\begin{figure}
\begin{center}
\leavevmode
\hbox{
\epsfxsize = 6.5in
\epsffile{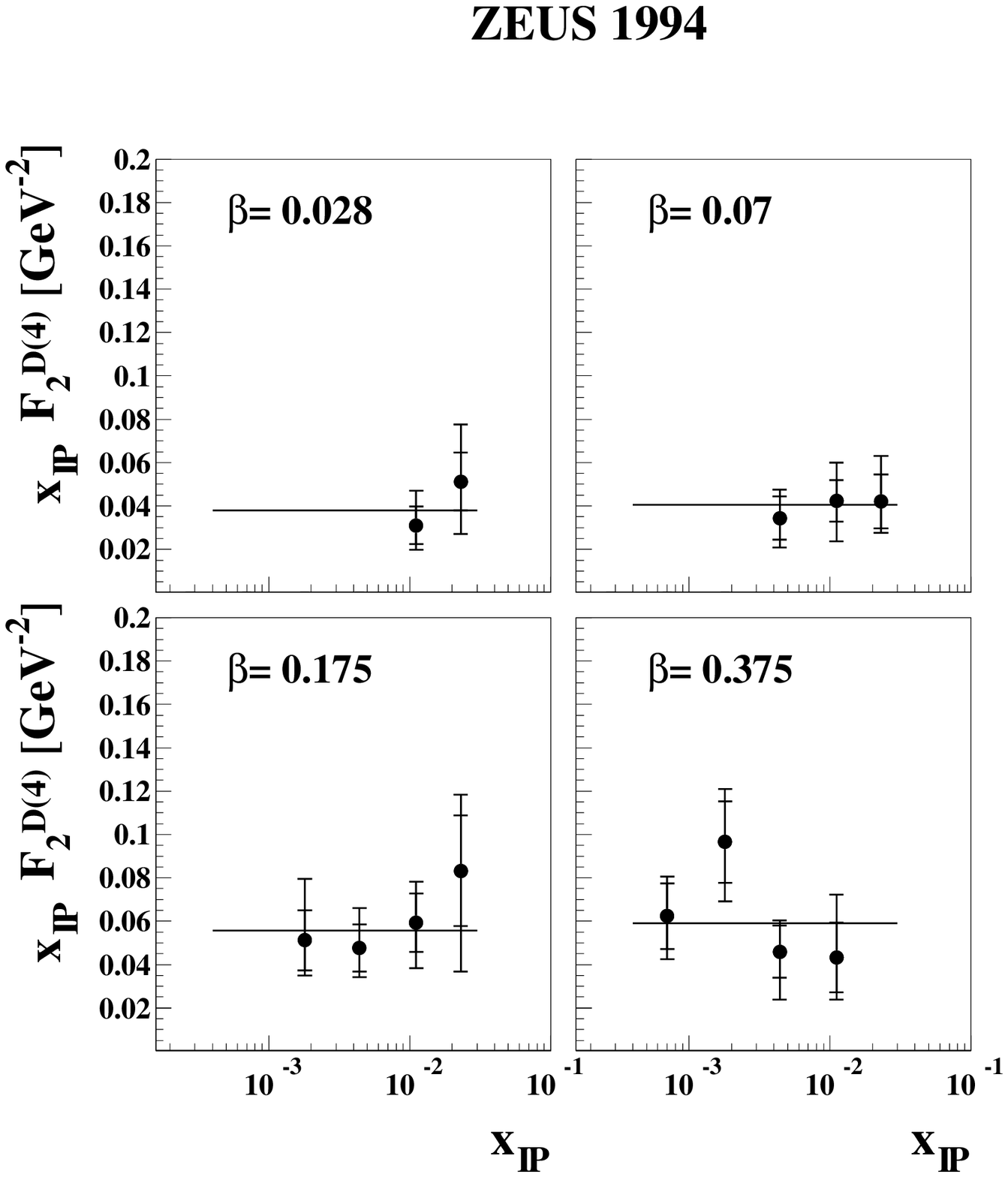}}
\end{center}
\caption{ 
The structure function $\xpom \cdot \ftwodfour(\beta,Q^2,\xpom,t)$, plotted as a function
of $\xpom$ in four $\beta$ bins, at $<Q^2>=8~\gevtwo$ and $<|t|>=0.17~\gevtwo$. 
The inner error bars indicate the statistical errors; the outer
error bars are the sum of the statistical and systematic errors added in quadrature.
The $5.5\%$ normalization uncertainty is not included. 
The solid line
corresponds to the fit described in the text.
}
\end{figure}

\clearpage
\begin{figure}
\begin{center}
\leavevmode
\hbox{
\epsfysize = 6.5in
\epsffile{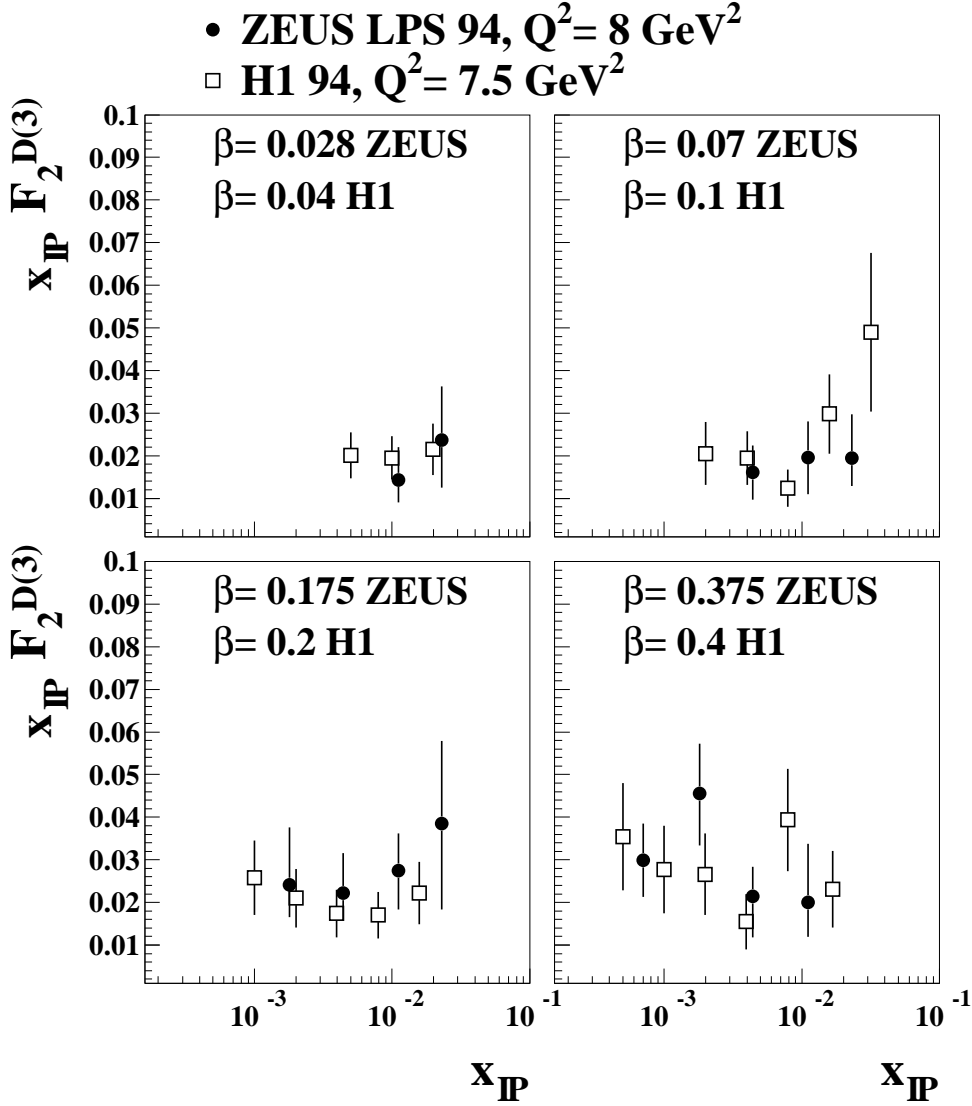}}
\end{center}
\caption{ 
The function $\xpom \cdot \ftwod(\beta,Q^2,\xpom)$ 
for this analysis at $<Q^2>=8~\gevtwo$ and $ 0< |t| < \infty$, compared to the results 
at $<Q^2>=7.5~\gevtwo$ and $|t|<1~\gevtwo$ from ref. [5]. 
The statistical and systematic errors are
added in quadrature. 
The $5.5\%$ and $6\%$ normalization uncertainties of the 
ZEUS 1994 LPS data and of the H1 1994 data, respectively, are
not included. 
}
\end{figure}

\clearpage
\begin{figure}
\begin{center}
\leavevmode
\hbox{
\epsfxsize = 6.5in
\epsffile{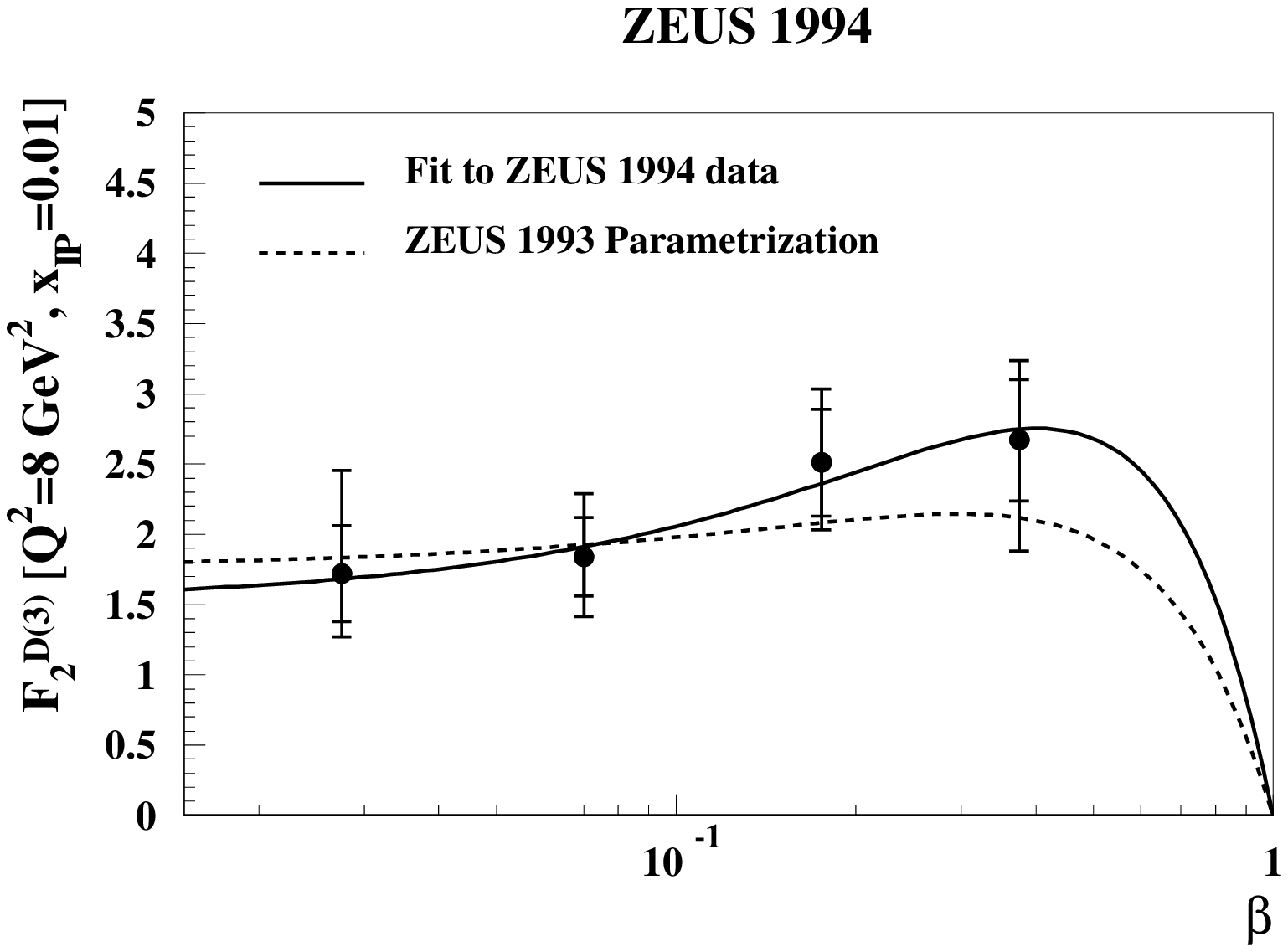}}
\end{center}
\vspace{-7cm}
\caption{ 
The value of $\ftwod(\beta,<Q^2>=8~\gevtwo,\xpom=0.01)$, obtained from the fits
to the individual bins in $\beta$. The solid line represents the fit
to the data as described in the text.
The dashed line indicates the parameterization of ref. [4]
scaled down to
remove the estimated $15\%$ double dissociative contributions.
}
\end{figure}

\end{document}